\documentclass[11pt,a4paper]{article}

\usepackage{jcappub}
\usepackage{graphicx}
\usepackage{amssymb}
\usepackage{amsmath}
\usepackage{bm}
\usepackage{latexsym}
\usepackage{epsfig}
\usepackage{textcomp}
\usepackage{color}

\makeatletter
\gdef\@fpheader{}
\makeatother

\def\nn{\nonumber} 
\def\pa{{\partial}}
\def\f{\frac}
\def\l{\left}
\def\r{\right}
\def\d{{\rm d}}
\def\Mpl{M_{_{\rm Pl}}}
\def\beq{\begin{equation}}
\def\eeq{\end{equation}} 
\def\beqa{\begin{eqnarray}}
\def\eeqa{\end{eqnarray}} 

\def\bA{\bar A}
\def\cA{\mathcal A}
\def\cN{\mathcal N}
\def\psb{{\mathcal P}_{_{\rm B}}}
\def\pse{{\mathcal P}_{_{\rm E}}}

\newcommand{\viz}{\textit{viz.~}}
\newcommand{\ie}{\textit{i.e.~}}

\title{Generation of scale invariant magnetic fields in bouncing universes}
\author[a]{L.~Sriramkumar,}
\author[b]{Kumar Atmjeet,}
\author[c]{and Rajeev Kumar Jain} 
\affiliation[a]{Department of Physics, Indian Institute of Technology Madras, 
Chennai~600036, India}
\affiliation[b]{Department of Physics and Astronomy, University of Delhi, 
Delhi~110007, India}
\affiliation[c]{CP$^3$-Origins, Centre for Cosmology and Particle Physics Phenomenology, \\
University of Southern Denmark, Campusvej 55, 
5230 Odense M, Denmark}
\emailAdd{sriram@physics.iitm.ac.in}
\emailAdd{katmjeet@physics.du.ac.in}
\emailAdd{jain@cp3.dias.sdu.dk}

\abstract{We consider the generation of primordial magnetic fields in 
a class of bouncing models when the electromagnetic action is coupled 
non-minimally to a scalar field that, say, drives the background 
evolution.
For scale factors that have the power law form at very early times and 
non-minimal couplings which are simple powers of the scale factor, one 
can easily show that scale invariant spectra for the magnetic field 
can arise {\it before the bounce}\/ for certain values of the indices 
involved.
It will be interesting to examine if these power spectra retain their
shape {\it after the bounce}.\/
However, analytical solutions for the Fourier modes of the electromagnetic vector 
potential across the bounce are difficult to obtain.
In this work, with the help of a new time variable that we introduce, which
we refer to as the ${\rm e}$-${\cal N}$-fold, we investigate these scenarios 
numerically.
Imposing the initial conditions on the modes in the contracting phase, we 
numerically evolve the modes across the bounce and evaluate the spectra of 
the electric and magnetic fields at a suitable time after the bounce. 
As one could have intuitively expected, though the complete spectra depend on 
the details of the bounce, we find that, under the original conditions, scale 
invariant spectra of the magnetic fields do arise for wavenumbers much smaller 
than the scale associated with the bounce.
We also show that magnetic fields which correspond to observed strengths  
today can be generated for specific values of the parameters.
But, we find that, at the bounce, the backreaction due to the electromagnetic
modes that have been generated can be significantly large calling into question 
the viability of the model. 
We briefly discuss the implications of our results.}
\arxivnumber{}

\begin{document}
\maketitle


\section{Introduction}

Over the last decade or so, a considerable amount of attention has been 
devoted in the literature to study models of the universe wherein, {\it as 
we go back in time},\/ the scale factor, rather than vanish, decreases to a 
small value, and starts to increase again (see, for instance,
Refs.~\cite{Finelli:2001sr,Peter:2002cn,Peter:2003rg,Martin:2003sf,Martin:2003bp,Allen:2004vz,Martin:2004pm,Creminelli:2004jg,Creminelli:2007aq,Cai:2007qw,Abramo:2007mp,Finelli:2007tr,Falciano:2008gt,Qiu:2011cy};
for reviews, see Refs.~\cite{Novello:2008ra,Battefeld:2014uga}).
In other words, in these `bouncing' scenarios, the universe goes through a 
contracting phase, decreases in size to a finite, but {\it non-zero}\/ value,
before it begins to expand. 
Interestingly, such a behavior for the scale factor helps in overcoming the 
horizon problem, without the need for an inflationary epoch\footnote{There will 
indeed be an epoch of accelerated expansion near the bounce as the universe 
transits from a contracting to an expanding phase. 
However, the period of acceleration need not last for $60$-$70$ e-folds, as it 
is required in the standard inflationary picture to resolve the horizon problem.}. 
One finds that, in such scenarios, well motivated, vacuum like, initial conditions 
can be imposed on the perturbations at early times during the contracting phase, as 
in the conventional inflationary scenarios~\cite{Novello:2008ra,Battefeld:2014uga}.

\par

Indeed, there exist (genuine) concerns whether such an assumption about the 
scale factor is valid in a domain where general relativity is expected to 
fail and quantum gravitational effects are supposed to take over. 
We shall ignore such concerns and refer the reader to the literature that 
discuss such issues~\cite{Novello:2008ra,Battefeld:2014uga}. 
Further, it is well known that matter fields may have to violate the null energy 
condition near the bounce in order to give rise to such a scale factor. 
In this work, we shall not attempt to construct models of matter fields that 
will give rise to a bounce, but shall simply assume such a behavior for the 
scale factor. 

\par

Various observations point to the presence of coherent magnetic fields in 
large scale cosmological structures as well as in the intergalactic 
medium (in this context, see Refs.~\cite{Grasso:2000wj,Widrow:2002ud}; for 
some recent reviews, see Refs.~\cite{Kandus:2010nw,Widrow:2011hs,Durrer:2013pga,Subramanian:2015lua}).
While the field strength in the galaxies and clusters is about a few micro-Gauss, 
an interesting lower bound of a few femto-Gauss on the coherent magnetic fields 
in the intergalactic medium has been  
obtained~\cite{Neronov:1900zz,Tavecchio:2010mk}.
These observations cannot be explained by astrophysical process alone and, it 
seems inevitable that, at least on the largest scales, the magnetic fields have 
a cosmological origin.
This has led to the construction of various mechanisms for the generation of 
magnetic fields in the early universe.
A lot of effort in this direction has focused on the production of magnetic 
fields during the inflationary phase~\cite{Turner:1987bw,Ratra:1991bn,Bamba:2003av,Bamba:2006ga,Martin:2007ue,Campanelli:2008kh,Demozzi:2009fu,Subramanian:2009fu,Kanno:2009ei,Durrer:2010mq,Urban:2011bu,Byrnes:2011aa,Jain:2012jy,Kahniashvili:2012vt,Cheng:2014kga}.
It has long been known that the conformal invariance of the electromagnetic 
field has to be broken if magnetic fields are to be produced in the early 
universe. 
Else, the expansion of the universe leads to an adiabatic dilution of the 
magnetic field strength to insignificant levels. 
Often, it is found that, in order to generate magnetic fields of sufficient
amplitude during inflation, the coupling function which breaks the conformal
invariance of the electromagnetic action has to grow rapidly at late times.
Due to this reason, the simplest scenarios of inflationary magnetogenesis have 
been shown to suffer from either the backreaction or the strong coupling 
problem~\cite{Demozzi:2009fu,Ferreira:2013sqa,Ferreira:2014hma}.

\par

In this work, we study the generation of primordial magnetic fields in bouncing
models.
We consider a class of scenarios wherein the electromagnetic field is coupled 
non-minimally to a scalar field, which can be possibly driving the bounce.
Clearly, it will be interesting to examine if scale invariant magnetic fields with 
sufficient strengths can be generated in such models~\cite{Salim:2006nw,Membiela:2013cea}. 
If one considers scale factors that behave as a power law at early times and
a coupling function which depends on a power of the scale factor, one can easily
argue that scale invariant power spectra for the magnetic field can arise before
the bounce for certain values of the indices involved. 
An interesting question to ask is if these power spectra will retain their shape
even after the bounce.
Since analytic solutions to the modes of the electromagnetic vector potential
across the bounce seem difficult to obtain, we investigate the problem numerically.
We begin by introducing a new time variable which allows for efficient numerical 
integration of the equation of motion governing the vector potential.
Imposing the initial conditions at sufficiently early times before the bounce, we 
evolve the Fourier modes across the bounce and evaluate the power spectra at a
suitable time after the bounce.
We illustrate that scale invariant magnetic fields are indeed generated under the 
original conditions provided the wavenumbers involved are much smaller than the 
characteristic scale associated with the bounce.

\par

This paper is organized as follows: 
In the next section, we shall quickly sketch a few essential aspects of 
non-minimally coupled electromagnetic fields.
We shall discuss the equation of motion governing the electromagnetic
vector potential, the quantization of the potential and the power 
spectra associated with the electric and magnetic fields.
In Sec.~\ref{sec:m}, we shall describe the form of the bounce and the 
type of non-minimal coupling that we shall consider.
In Sec.~\ref{sec:ps}, we shall outline the numerical method that we adopt
to integrate the equation of motion and arrive at the power spectra 
describing the electric and magnetic fields.
We shall illustrate that strictly scale invariant magnetic fields arise 
for a class of couplings and bouncing scenarios.
In Sec.~\ref{sec:os}, we shall examine if magnetic fields that correspond 
to observable strengths today can be generated in the bouncing models.
In Sec.~\ref{sec:br}, we shall discuss the issue of backreaction in these
situations.
Finally, we shall close with Sec.~\ref{sec:d} with a summary of our results 
and a brief outlook.

\par

Before we proceed further, a few words on the conventions and notations that 
we shall adopt are in order. 
We shall work with natural units such that $\hbar=c=1$, and set the Planck mass 
to be $\Mpl \equiv (8\, \pi\, G)^{-1/2}$.
We shall adopt the metric signature of $(-,+,+,+)$. 
Note that Greek indices shall, in general, denote the spacetime coordinates, 
modulo $\lambda$ which we shall use to denote the polarization of the
electromagnetic field.
Similarly, Latin indices, apart from $k$ which shall be reserved for denoting 
the wavenumber, shall represent the spatial coordinates.
Lastly, an overprime shall denote differentiation with respect to the conformal 
time coordinate.


\section{The non-minimal action, equations of motion and 
power spectra}\label{sec:nma}

We shall consider a case wherein the electromagnetic field is coupled 
non-minimally to a scalar field $\phi$ and is described by 
the action (see, for instance, 
Refs.~\cite{Bamba:2006ga,Martin:2007ue,Subramanian:2009fu})
\begin{equation}\label{eq:nm-em}
S[\phi,A^{\mu}]=-\frac{1}{16\,\pi} \int \d^{4}x\, \sqrt{-g}\, J^2(\phi)\, 
F_{\mu\nu}F^{\mu\nu},
\end{equation}
where $F_{\mu\nu}$ denotes the electromagnetic field tensor which is given in 
terms of the vector potential $A^{\mu}$ as follows:
\begin{equation}
F_{\mu\nu} = A_{\nu;\mu}-A_{\mu;\nu}= A_{\nu,\mu}-A_{\mu,\nu}.
\end{equation}
The scalar field $\phi$ could be, for instance, the primary matter field that 
is driving the background evolution and $J$ is an arbitrary function of the 
field.
As we had mentioned earlier, in order to generate primordial magnetic fields, 
it is necessary to break the conformal invariance of the electromagnetic field.
Although there exist a variety of possibilities for breaking the conformal 
invariance, the above non-minimal action is one of the simplest gauge-invariant 
choices. 
Evidently, it is the coupling function $J$ that is responsible for breaking the 
conformal invariance of the action.

\par

Variation of the above action leads to the following equation of motion describing 
the evolution of the electromagnetic field:
\begin{equation}
\frac{1}{\sqrt{-g}}\;{\partial}_{\mu} \l[\sqrt{-g}\, J^2(\phi)\, F^{\mu\nu}\r]=0.
\end{equation}  
Consider a $(3 + 1)$-dimensional, spatially flat, 
Friedmann-Lema\^{\i}tre-Robertson-Walker (FLRW) universe described by 
line-element
\begin{equation}
\d s^2 = -\d t^2 + a^{2}(t)\, \d{\bm x}^2 
= a^{2}(\eta)\, \l(-\d\eta^{2} + \d{\bm x}^2\r),\label{eq:frwle}
\end{equation}
where $t$ is the cosmic time, $a(t)$ is the scale factor and $\eta=\int 
\d t/a(t)$ denotes the conformal time coordinate. 
Let us choose to work in the Coulomb gauge wherein $A_{0} = 0$ and 
$\pa_{i}A^{i}=0$. 
In such a gauge, the equation governing the spatial components of the vector
potential is found to be 
\begin{equation}
A_i''+2\,\frac{J'}{J}\,A_i'-a^{2}\,\pa_j\,\pa^j A_i=0.
\end{equation}

\par

In the Coulomb gauge, upon quantization, the vector potential ${\hat A}_i$ 
can be Fourier decomposed as follows (see, for instance, 
Refs.~\cite{Martin:2007ue,Subramanian:2009fu,Ferreira:2013sqa}):
\begin{equation}
{\hat A}_{i}(\eta,\mathbf{x})
=\sqrt{4\,\pi}\,\int{\frac{\d^3{\bm k}}{(2\,\pi)^{3/2}}
\sum_{\lambda=1}^2 {\tilde \epsilon}_{\lambda\,i}({\bm k})\,
\left[{\hat b}^{\lambda}_{\bm k}\, 
{\bar A}_{k}(\eta)\, {\rm e}^{i\,{\bm k}\cdot{\bm x}}
+{\hat b}^{\lambda}_{\bm k}{}^{\dag}\, {\bar A}_k^{*}(\eta)\,
{\rm e}^{-i\,{\bm k}\cdot{\bm x}}\right]},
\end{equation}
where the modes ${\bar A}_{k}$ satisfy the differential equation
\begin{equation}
{\bar A}_k''+2\,\f{J'}{J}\,{\bar A}_k'+k^2\, {\bar A}_k=0. 
\label{eq:de-Abk-o}
\end{equation}
In the above Fourier decomposition of the vector potential, the summation 
over $\lambda$ corresponds to the two orthonormal transverse polarizations.
The quantities ${\tilde \epsilon}^{i}_{\lambda}$ represent the polarization 
vectors, which form a part of the following orthonormal set of basis four 
vectors:
\begin{equation}
\epsilon_0^{\mu}=\l(\f{1}{a},{\bm 0}\r), \quad
\epsilon_\lambda^{\mu}
=\l(0,\f{{\tilde \epsilon}^{i}_{\lambda}}{a}\r)\quad{\rm and}\quad
\epsilon_3^{\mu}=\l(0,\f{k^i}{k\,a}\r).
\end{equation}
The three vectors ${\tilde \epsilon}^{i}_{\lambda}$ are unit vectors that 
are orthogonal to each other and to the wavevector~${\bm k}$.
Hence, they satisfy the conditions
$\delta_{ij}\,
{\tilde \epsilon}^{i}_{\lambda}\,{\tilde \epsilon}^{j}_{\lambda}=1$
(with no summation over $\lambda$) and $\delta_{ij}\,k^i\, 
{\tilde \epsilon}^{j}_{\lambda}=0$.
Moreover, the operators ${\hat b}^{\lambda}_{\bm k}$ and 
${\hat b}^{\lambda}_{\bm k}{}^{\dag}$ are the annihilation and creation 
operators which satisfy the standard commutation relations, \viz
\begin{equation}
[{\hat b}^{\lambda}_{\bm k},{\hat b}^{\lambda'}_{\bm k'}]
=[{\hat b}^{\lambda}_{\bm k}{}^{\dag},{\hat b}^{\lambda'}_{\bm k'}{}^{\dag}]
=0
\quad{\rm and}\quad
[{\hat b}^{\lambda}_{\bm k},{\hat b}^{\lambda'}_{\bm k'}{}^{\dag}]
=\delta_{\lambda \lambda'}\;\delta^{(3)}\l({\bm k}-{\bm k'}\r). 
\end{equation}
If we further define a new variable ${\mathcal A}_k=J\, {\bar A}_k$, then 
the equation~(\ref{eq:de-Abk-o}) for ${\bar A}_k$ simplifies to 
\begin{equation}
{\mathcal A}_k''+\l(k^2-\f{J''}{J}\r)\,{\mathcal A}_k=0.\label{eq:de-cAk}
\end{equation}

\par

One can show that the energy densities associated with the electric and magnetic 
fields can be written in terms of the vector potential $A_i$ and its time and 
spatial derivatives as follows:
\begin{eqnarray}
\rho_{_{\rm E}} &=&\f{J^2}{8\,\pi\,a^2}\; g^{ij}\, A_i'\, A_j',\\
\rho_{_{\rm B}} &=&\f{J^2}{16\,\pi}\; g^{ij}\, g^{lm}\,
\l(\pa_j A_m-\pa_m A_j\r)\,\l(\pa_i A_l-\pa_l A_i\r),
\end{eqnarray}
where $g^{ij}=\delta^{ij}/a^2$ denotes the spatial components of the FLRW metric.
The expectation values of the corresponding operators, \ie ${\hat \rho}_{_{\rm E}}$
and ${\hat \rho}_{_{\rm B}}$, can be evaluated in the vacuum state annihilated by 
the operator ${\hat b}^{\lambda}_{\bm k}$.
It can be shown that the spectral energy densities of the magnetic and electric 
fields are given by~\cite{Martin:2007ue,Subramanian:2009fu}
\begin{subequations}
\label{eq:ps}
\begin{eqnarray}
\psb(k)&=&\f{\d\langle 0\vert {\hat \rho}_{_{\rm B}}\vert 0\rangle}{\d\ln k}
=\f{J^2(\eta)}{2\,\pi^{2}}\,\f{k^{5}}{a^{4}(\eta)}\,
\vert{\bar A}_k(\eta)\vert^{2},\label{eq:psb}\\
\pse(k)&=&\f{\d\langle 0\vert {\hat \rho}_{_{\rm E}}\vert 0\rangle}{\d\ln k}
=\f{J^{2}(\eta)}{2\,\pi^{2}}\,\frac{k^{3}}{a^{4}(\eta)}\,
\vert{\bar A}_k'(\eta)\vert^2.\label{eq:pse}
\end{eqnarray}
\end{subequations}
These spectral energy densities are often referred to as the power spectra for 
the generated magnetic and electric fields respectively. 
A flat or scale invariant magnetic field spectrum corresponds to a constant, 
\ie\/ $k$-independent, ${\mathcal P}_{_{\rm B}}(k)$.


\section{Modeling the bounce and the non-minimal coupling term}\label{sec:m}
 
We shall model the bounce by assuming that the scale factor $a(\eta)$ 
behaves as follows:
\begin{equation}
a(\eta)=a_{0}\, \l[1+\l(\eta/\eta_{0}\r)^{2}\r]^{p},\label{eq:sf}
\end{equation}
where $a_{0}$ is the value of the scale factor at the bounce (\ie\/ 
when $\eta=0$), $\eta_{0}$ denotes the time scale that determines the 
duration of the bounce and $p>0$. 
Such a simple choice for the scale factor leads to a symmetric non-singular 
bounce for which the Hubble parameter $H=a'/a^2$ vanishes at the bounce.   
In the absence of any detailed modeling, it is natural to expect that 
$\eta_{0}$ is determined by the fundamental Planck scale, \ie 
$\eta_0^{-1} \simeq \Mpl$. 
As we shall point out in due course, wavenumbers of cosmological interest
are considerably smaller than the scale $k_0=\eta_0^{-1}$.
Therefore, as will be clear from our discussion, our essential conclusion, 
viz. that bouncing models can lead to scale invariant spectra for the magnetic
field over cosmological scales, remains unaffected even if one chooses to works 
with an $\eta_0^{-1}$ that is a few orders of magnitude below $\Mpl$.
It is useful to note that the above scale factor reduces to the simple 
power law form with $a(\eta)\propto \eta^{2\,p}$ at very early times
(\ie\/ when $-\eta\gg \eta_0$).

\par

We should stress that the condition $p>0$ ensures that the universe does 
not go through an accelerated contraction during its early stages (\ie as 
$\eta\to -\infty$). 
This, in turn, will ensure that, at these early times, the modes of 
electromagnetic vector potential are well inside the Hubble radius, 
thereby allowing us to impose the standard, sub-Hubble, Bunch-Davies 
initial condition.
Actually, it is not essential that the modes of the vector potential 
should be inside the Hubble radius in order for us to be able to 
impose Minkowski-like initial conditions during the contracting phase. 
It would suffice if the conformal invariance of the electromagnetic 
field, which we shall assume to be broken around the bounce, is 
restored at a sufficiently early epoch. 
In fact, even this is a strong demand and it is not necessary. 
As the equation~(\ref{eq:de-cAk}) suggests, we can impose well motivated 
Minkowski-like initial conditions at a suitably early time wherein 
$k^2\gg J''/J$.
But, we shall require the modes to be inside the Hubble radius at early 
times, if we expect the scalar perturbations too to be seeded during the 
contracting phase. 

\par

Note that the coupling function $J$ is actually assumed to be a function 
of the scalar field $\phi$ that drives the background evolution.
As we have discussed, in this work, rather than attempt to model the 
bounce, we shall assume a specific form for the scale factor
[viz. Eq.~(\ref{eq:sf})].
Due to this reason, we shall also assume a particular form for the 
coupling function~$J$.
It seems natural to expect that the coupling function will either grow
or decay away from the bounce.
In order to describe such behavior, we shall conveniently express the
coupling function in terms of the scale factor as follows:
\begin{equation}
J(\eta)=J_0\, a^n(\eta).\label{eq:J}
\end{equation}
Clearly, for $n>0$ the minimum of $J(\eta)$ is located at the bounce, while 
it grows away from the bounce.
In contrast, for $n<0$, $J(\eta)$ is maximum at the bounce and decays away
from it.
In the context of power law inflation, for the above coupling function, it 
is known that specific choices for the indices lead to scale invariant 
spectra for the magnetic field~\cite{Martin:2007ue,Subramanian:2009fu}.
As we shall show later, similar choices for $p$ and $n$ lead to scale 
invariant magnetic fields over cosmological scales in the bouncing 
scenario too.


\section{Numerical analysis and results}\label{sec:ps}

In this section, we shall first outline the procedure that we shall adopt 
to numerically integrate the equation of motion governing the electromagnetic
vector potential and then present the results we obtain for the class of 
bouncing scenarios described above.


\subsection{The numerical procedure}

Let us now sketch the numerical procedure that we shall follow to solve 
the equation of motion~(\ref{eq:de-Abk-o}) governing the modes $\bA_k$
of the electromagnetic vector potential.


\subsubsection{E-$\cN$-folds as the time variable}

Our first challenge is to identify a convenient independent variable for 
integrating the differential equation.
At first look, this may not seem to be an important issue, but a suitable 
choice can obviously make numerical integration stable, quick and easy 
to handle.
Let us explain.
It is well known that it is rather inefficient to integrate the equations 
of motion governing fields and perturbations in a FLRW universe in terms 
of either the cosmic or the conformal time coordinates.
Because of this reason, for instance, in the inflationary scenario, while
integrating numerically, one always uses the e-fold as the independent 
variable that represents time (in this context, see, for instance, 
Refs.~\cite{Salopek:1988qh,Ringeval:2007am,Jain:2008dw,Jain:2009pm,Hazra:2012yn}).
Recall that, the conventional e-fold $N$ is defined $N={\rm log}\,(a/a_0)$ so 
that $a(N)=a_0\, {\rm exp}\,N$, where, evidently, $N=0$ is a suitable time at
which the scale factor takes the value $a_0$. 
Due to the exponential function involved, a relatively small range in e-folds 
covers a wide range in time and scale factor.

\par 

However, the function ${\rm e}^N$ is a monotonically increasing function 
of $N$. 
As a result, while e-folds are useful in characterizing expanding universes, 
it does not seem to be a suitable time variable to describe bouncing 
models\footnote{One can possibly work with functions ${\rm e}^N$ after the 
bounce and ${\rm e}^{-N}$ prior to the bounce to describe the scale factor.
But, the resulting cusp at $N=0$ is highly undesirable.}. 
In a bouncing scenario, it seems to be convenient to choose a suitable variable
to be zero at the bounce, with negative values corresponding to the contracting 
phase and positive values characterizing the expanding regime.
Further, since we are focusing on symmetric bounces, it seems natural to 
demand that the scale factor is symmetric in terms of the new independent
variable to characterize time.   
An obvious choice for the scale factor seems to be $a({\mathcal N})
=a_0\, {\rm exp}\, ({\mathcal N}^2/2)$, with ${\mathcal N}$ being the 
new time variable that we shall consider for integrating the 
differential equation~(\ref{eq:de-Abk-o}).
(The factor of two  has been introduced in the exponential for convenience.) 
For want of a better name, we shall refer to the variable ${\mathcal N}$ as 
e-$\cN$-fold since the scale factor grows roughly by the amount ${\rm e}^{\cN}$
between $\cN$ and $(\cN+1)$\footnote{To be precise, it has to be called 
e-$\cN+\f{1}{2}$-folds.
But, since ${\rm e}^{1/2}\simeq 1$, for convenience, we shall simply refer
to it as e-$\cN$-folds.}. 
Needless to add, because of the rapidly growing exponential involved, a wide 
range in scale factor can be covered by a relatively short span of e-$\cN$-folds.   


\subsubsection{Imposing the initial conditions and evaluating the power spectra}

In terms of the e-$\cN$-fold, the differential equation~(\ref{eq:de-Abk-o}) 
satisfied by $\bA_k$ can be expressed as
\begin{equation}
\f{\d^{2}\bA_k}{\d\cN^2}+\l(\f{1}{H}\, \f{\d H}{\d\cN}
+\f{2}{J}\, \f{\d J}{\d\cN}+\cN-\f{1}{\cN}\r)\, \f{\d \bA_k}{\d\cN}
+\l(\f{k\, \cN}{a\, H}\r)^2\, \bA_k=0.\label{eq:de-Abk} 
\end{equation}
Since the scale factor and the non-minimal coupling function $J$ are specified
[cf. Eqs.~(\ref{eq:sf}) and~(\ref{eq:J})], evidently, the coefficients of the 
above differential equation are straightforward to determine.
So, if the initial conditions are provided, the differential equation can be
integrated numerically using the standard methods.
We shall discuss about the initial conditions below. 
Here, we wish to make a clarifying remark regarding arriving at the Hubble 
parameter in terms of the e-$\cN$-fold.
Note that the form of the scale factor is specified in terms of the conformal
time coordinate [cf. Eq.~(\ref{eq:sf})].
The corresponding Hubble parameter $H$ can be easily evaluated in terms
of the conformal time $\eta$. 
In order to arrive at the expression for the Hubble parameter in terms of 
e-$\cN$-fold, we shall require the relation between $\eta$ and $\cN$.
This can be obtained from the fact that $a(\cN)=a_0\,{\rm e}^{\cN^2/2}$ and 
the expression~(\ref{eq:sf}) for the scale factor.
However, caution should be exercised in arriving at the relation. 
Since the Hubble parameter is negative during the contracting phase and positive 
in the expanding regime, one has to choose the suitable root for $\eta(\cN)$ in 
order to describe $H$ correctly.
For the scale factor~(\ref{eq:sf}), we find that the function $\eta(\cN)$ is given 
by 
\begin{equation}
\eta(\cN)=\pm\,\eta_0\, \l\{\l[a(\cN)/a_0\r]^{1/p}-1\r\}^{1/2}
=\pm\, \eta_0\, \l[{\rm e}^{\cN^2/(2\,p)}-1\r]^{1/2},
\end{equation}
with the negative root corresponding to the period before the bounce and the 
positive root after.
\par 

Let us now turn to the initial conditions.
During inflation, while imposing the standard initial conditions on either 
the scalar or the tensor perturbations, one requires that the modes are 
sufficiently deep inside the Hubble radius.
As we have already discussed, such a demand is not necessary in the case of the 
electromagnetic modes.
Since the FLRW universe is conformally flat, Minkowski-like initial conditions 
can be imposed on the modes $\cA_k$ in a domain wherein the non-minimal coupling 
reduces to unity.
In fact, even such a requirement is not essential and it is sufficient if there
exists an early time during the contracting phase when 
$k^2\gg J''/J$\footnote{Actually, a similar condition during the inflationary
phase would suffice too. 
But, during inflation, such a condition corresponds to the condition for the 
modes to be inside the Hubble radius.}.
The standard initial conditions, \viz\/ that $\cA_k =1/\sqrt{2\,k}$ and $\cA_k'
=-i\,\sqrt{k/2}$, can be imposed at such an initial time.
Note that these initial conditions transform to the following conditions on
the variable $\bA_k$ and its derivative with respect to the e-$\cN$-fold:
\begin{subequations}
\label{eq:ic}
\begin{eqnarray}
\bA_k&=&\f{1}{J(\cN_i)}\,  \f{1}{\sqrt{2\,k}},\\
\f{\d\bA_k}{\d \cN}
&=&-\f{i\ \cN_i}{a(\cN_i)\,H(\cN_i)\,J(\cN_i)}\,\sqrt{\f{k}{2}} 
-\f{1}{J^2(\cN_i)}\,\f{\d J(\cN_i)}{\d\cN}\,\f{1}{\sqrt{2\,k}},
\end{eqnarray}
\end{subequations}
where $\cN_i$ denotes the e-$\cN$-fold when the initial conditions are imposed.

\par

\begin{figure}[!t]
\begin{center}
\includegraphics[width=7.65cm]{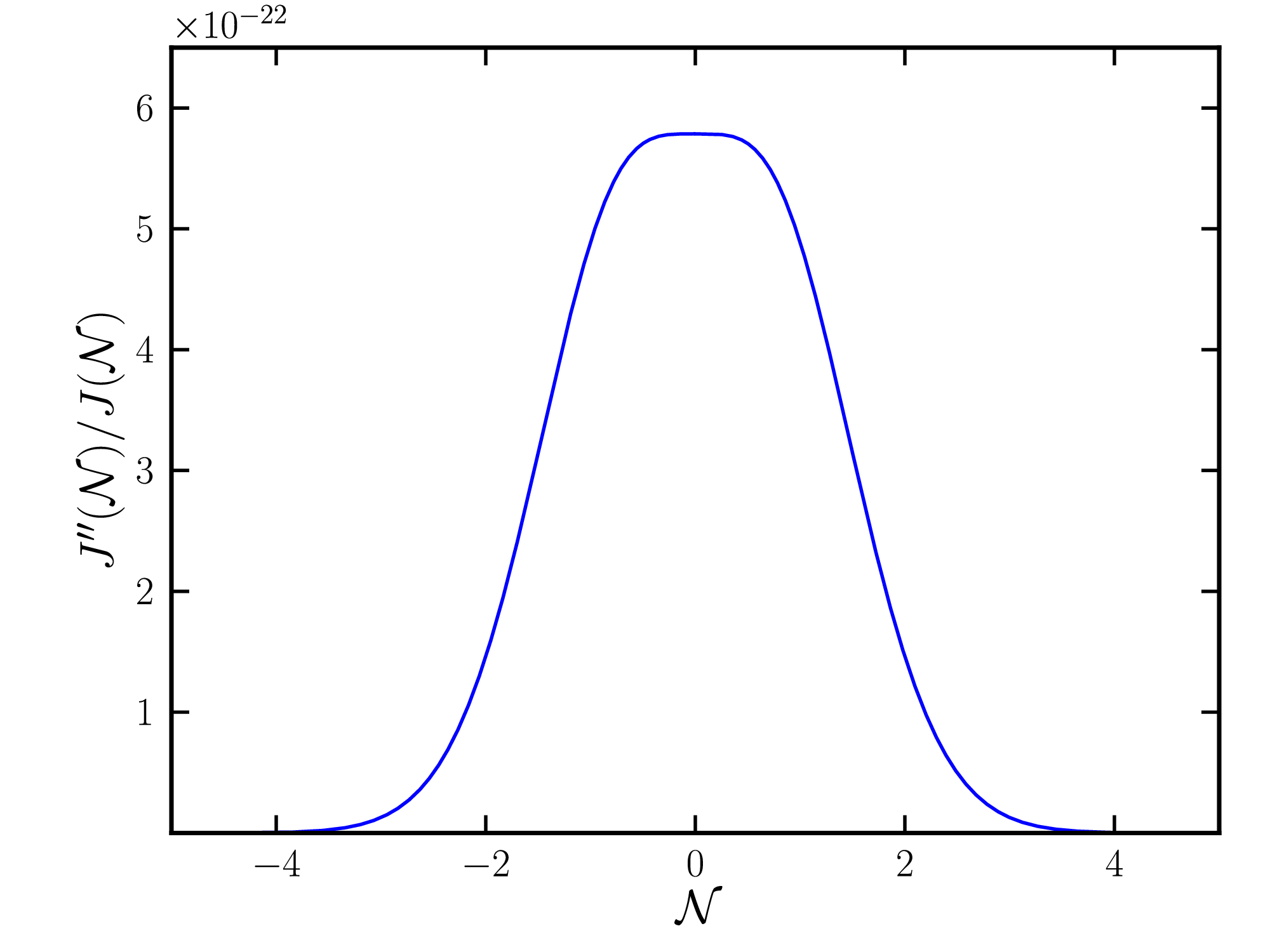}
\includegraphics[width=7.65cm]{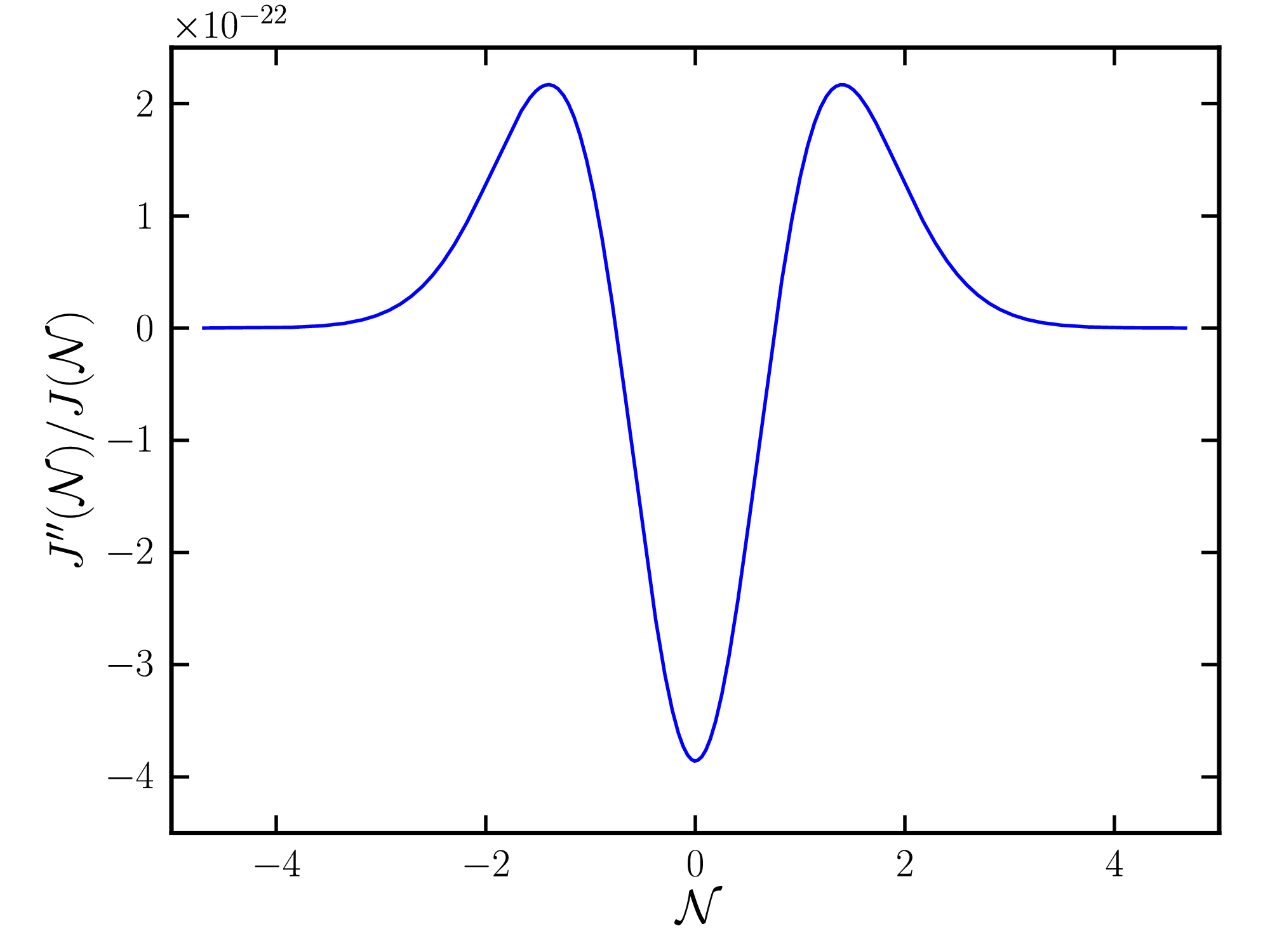}
\end{center}
\caption{The behavior of the quantity $J''/J$ has been plotted as a function
of $\cN$ for $p=1$ and $n=3/2$ (on the left) and $n=-1$ (on the right).
As we shall see, these choices for the indices $p$ and $n$ lead to scale
invariant spectra for the magnetic field.
The quantity that we have plotted is independent of $J_0$ and $a_0$, and we 
have chosen $\eta_0={\rm e}^{25}=7.2\times10^{10}$.
We should mention that this choice for $\eta_0$ is arbitrary and it has been 
chosen for illustrative purposes.
We shall work with a more plausible value of $\eta_0$ in due course.
Note that the maximum value of $J''/J$ is roughly of the order of $1/\eta_0^2$.
Evidently, while $J''/J$ has a single maximum when $n=3/2$, it has two maxima
when $n=-1$.}
\label{fig:JppbJ}
\end{figure}

Numerically, we impose the initial conditions at the earliest time when the 
condition $k^2=100\,(J''/J)$ is satisfied.
In Fig.~\ref{fig:JppbJ}, we have plotted the quantity $J''/J$ for certain 
values of the parameters involved.
Note that, while $J''/J$ exhibits a single maximum when $J$ is growing away
from the bounce, it has two maxima when $J$ decays. 
We should stress here that, in the latter case, to impose the initial 
conditions, we need to carefully choose the earliest time wherein the 
condition $k^2=100\,(J''/J)$ is satisfied.
Moreover, it should be pointed out that the initial conditions are imposed 
at different e-$\cN$-folds for different $k$. 
Further, since the maximum value of $J''/J$ is of the order of $1/\eta_0^2=k_0^2$, 
such a condition will always be satisfied by modes with wavenumbers such that 
$k\lesssim k_0$.
Actually, we find that modes with wavenumbers upto $k\simeq {\cal O}(10)\,k_0$ 
satisfy the above condition.
We evaluate the power spectra for modes with wavenumbers up till these values.

\par

The last point that we need to discuss concerns the evaluation of the power 
spectra.
After having imposed the initial conditions at an early time prior to the bounce, 
we evolve these modes across the bounce and compute the power spectra $\psb(k)$
and $\pse(k)$ in the expanding phase.
We evaluate the spectra associated with all the modes at the specific time when 
the {\it smallest scale of interest}\/ satisfies the condition $k^2=10^3\,(J''/J)$.
Again, in the cases wherein $J$ decays, one needs to be careful in ensuring that 
the time is the latest time when the condition is satisfied.
We should mention here that this condition essentially corresponds
to choosing a domain soon after the bounce when the amplitude of the electromagnetic 
modes are largely constant for modes such that $k\ll k_0$.


\subsection{Results}

Before we go on to discuss the numerical results, let us try to understand
a few points analytically.
As we had mentioned earlier, at very early times (\ie\/ for $-\eta\gg \eta_0$),
the scale factor~(\ref{eq:sf}) simplifies to the power law form $a(\eta)\propto
\eta^{2\,p}$.
During such times, the non-minimal coupling function $J$ also has a power law 
form and it behaves as $J(\eta) \propto \eta^\alpha$, where we have set $\alpha=
2\,n\,p$.
In such a case, we have $J''/J\simeq \alpha\,(\alpha-1)/\eta^2$ and it is easy
to show that, for $k\ll k_0$, the solutions to the modes of the electromagnetic 
vector potential $\bA_k$ can be expressed in terms of the Bessel functions
$J_\nu(x)$, exactly as it occurs for a similar coupling function in power law 
inflation~\cite{Martin:2007ue,Subramanian:2009fu,Ferreira:2013sqa,Membiela:2013cea}.
One finds that the solutions can be expressed in terms of the quantity $\cA_k$ as 
follows:
\begin{equation}
\cA_k(\eta)  
= \sqrt{-k\, \eta}\, \l[C_{1}(k)\, J_{\alpha-1/2}(-k\, \eta) 
+ C_{2}(k)\, J_{-\alpha + 1/2} (-k\,\eta)\r],
\end{equation}
where the coefficients $C_{1}(k)$ and $C_{2}(k)$ are to be fixed by the initial 
conditions. 
Upon imposing the Bunch-Davies initial conditions as $k\,\eta\to-\infty$, one obtains 
that 
\begin{eqnarray}
C_1(k)&=&\sqrt{\f{\pi}{4\,k}}\; \f{{\rm e}^{-i\,\pi\,\alpha/2}}{{\rm cos}\l(\pi\,\alpha\r)},\\
C_2(k)&=&\sqrt{\f{\pi}{4\,k}}\; \f{{\rm e}^{i\,\pi\,(\alpha+1)/2}}{{\rm cos}\l(\pi\,\alpha\r)}.
\end{eqnarray}
It can also be shown that 
\begin{equation}
\cA_k'(\eta)-\f{J'}{J}\,\cA_k(\eta)   
= k\,\sqrt{-k\, \eta}\, \l[C_{1}(k)\, J_{\alpha+1/2}(-k\, \eta) 
- C_{2}(k)\, J_{-\alpha - 1/2} (-k\,\eta)\r].
\end{equation}
The spectra of magnetic and electric fields as one approaches the bounce, i.e. as $k\,\eta
\to 0^{-}$, can be arrived at from the above expressions for $\cA_k$ and  $\cA_k'-(J'/J)\cA_k$
and the asymptotic forms of the Bessel functions.
The magnetic field spectrum can be written 
as~\cite{Martin:2007ue,Subramanian:2009fu,Ferreira:2013sqa,Membiela:2013cea}
\begin{equation}
{\cal P}_{_{\rm B}}(k) 
= \frac{{\cal F}(m)}{2\,\pi^2}\, \l(\f{H}{2\,p}\r)^4 (-k\,\eta)^{4+2\,m},
\end{equation}
where $H\simeq (2\,p/a_0\,\eta)\, (\eta_0/\eta)^{2p}$, while $m=\alpha$ for $\alpha \le 1/2$ 
and $m =1-\alpha$ for $\alpha \ge 1/2$, 
Also, the quantity ${\cal F}(m)$ is given by
\begin{equation}
{\cal F}(m) = \frac{\pi}{2^{2\,m+1}\,\Gamma^2(m+1/2)\,\cos^2(\pi\,m)}.
\end{equation}
It is evident that $m=-2$ leads to a scale invariant spectrum for the magnetic field
which corresponds to either $\alpha=3$ or $\alpha=-2$.
The corresponding spectrum for the electric field can be obtained to be
\begin{equation}
{\cal P}_{_{\rm E}}(k) 
= \frac{{\cal G}(m)}{2\,\pi^2}\, \l(\f{H}{2\,p}\r)^4 (-k\,\eta)^{4+2\,m},
\end{equation}
where $m=1+\alpha$ if $\alpha \le -1/2$ and $m =-\alpha$ for $\alpha \ge -1/2$, while
${\cal G}(m)$ is given by
\begin{equation}
{\cal G}(m) = \frac{\pi}{2^{2\,m+3}\, \Gamma^2(m +3/2)\,\cos^2(\pi\,m)}. 
\end{equation}
Note that when, $\alpha=3$ and $\alpha=-2$, the power spectrum of the electric field 
behaves as $k^{-2}$ and $k^2$, respectively.
This implies that, in the bouncing scenario, one can expect these cases to lead 
to scale invariant spectra (over modes such that $k\ll k_0$) for the magnetic 
field before the bounce.
The question of interest is whether these spectra will retain their shape 
{\it after}\/ the bounce.
It is for this reason that we shall consider the combination of the indices $p$ 
and $n$ that correspond to either $\alpha=3$ or $-2$.

\par

It is also possible to analytically understand the behavior of the mode
$\bA_k$ and its time derivative $\bA_k'$ in the case wherein $n$ is 
positive.
Note that, when $n>0$, $J''/J$ has a maximum at the bounce.
In such a case, for $k\ll k_0$, $k^2\ll J''/J$ around the bounce.
From equation~(\ref{eq:de-Abk-o}), it is clear that in a domain where $k^2$
can be neglected, we have 
\begin{equation}
\bA_k''+2\,\f{J'}{J}\, \bA_k'\simeq 0,
\end{equation}
which can be integrated to yield
\begin{equation}
\bA_k'(\eta)\simeq\bA_k'(\eta_\ast)\,\f{J^2(\eta_\ast)}{J^2(\eta)},
\label{eq:Abpk-ar}
\end{equation}
where $\eta_\ast$ is a time when $k^2\ll J''/J$ before the bounce.
The above equation can be integrated to arrive at
\begin{eqnarray}
\bA_k(\eta)
\simeq \bA_k(\eta_\ast) +\bA_k'(\eta_\ast)\,\int\limits_{\eta_\ast}^{\eta}\d\eta\, 
\f{J^2(\eta_\ast)}{J^2(\eta)}
= \bA_k(\eta_\ast)
+\bA_k'(\eta_\ast)\,a^{2\,n}(\eta_\ast)\,
\int\limits_{\eta_\ast}^{\eta}\f{\d\eta}{a^{2\,n}(\eta)},
\end{eqnarray}
where we have set the constant of integration to be $\bA_k(\eta_\ast)$.
When $\alpha=3$, for the scale factor~(\ref{eq:sf}), we can evaluate the 
above integral to obtain that
\begin{eqnarray}
\bA_k(\eta)
&\simeq&\bA_k(\eta_\ast)+\bA_k'(\eta_\ast)\,\f{a^{2\,n}(\eta_\ast)}{a_0^{2\,n}}\,
\f{\eta_0}{8}\nn\\
& &\times\,\biggl\{\f{\eta}{\eta_0}\,\f{5+3\,(\eta/\eta_0)^2}{\l[1+
(\eta/\eta_0)^2\r]^2}+3\, {\rm tan}^{-1}\l(\f{\eta}{\eta_0}\r)
-\f{\eta_\ast}{\eta_0}\,\f{5+3\,(\eta_\ast/\eta_0)^2}{\l[1+
(\eta_\ast/\eta_0)^2\r]^2}-3\, {\rm tan}^{-1}\l(\f{\eta_\ast}{\eta_0}\r)\biggr\}.
\qquad\;\;\;\label{eq:Abk-ar}
\end{eqnarray}
We should point out that, while the first term is evidently a constant, the 
second term grows rather rapidly during the contracting phase and less so 
during the expanding regime.
We should mention that the above solutions for $\bA_k$ and $\bA_k'$ are 
valid around the bounce until the condition $k^2\ll J''/J$ is satisfied. 
In what follows, apart from presenting the numerical results for a few different
cases of interest, we shall also compare the above analytical results with the 
numerical ones for $\alpha=3$.

\par 

\begin{figure}[!h]
\begin{center}
\vskip -10pt
\includegraphics[width=6.0cm]{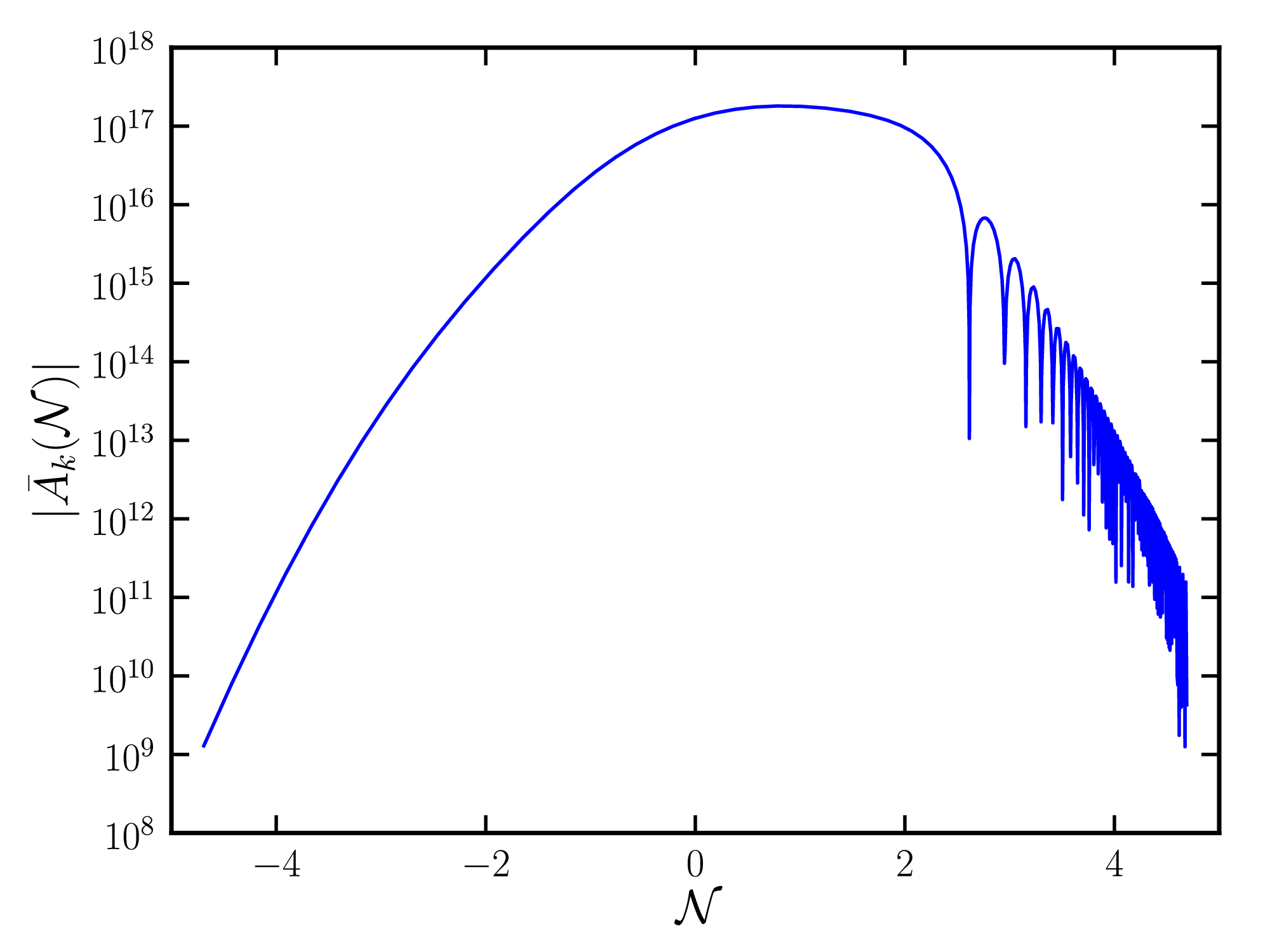}
\includegraphics[width=6.0cm]{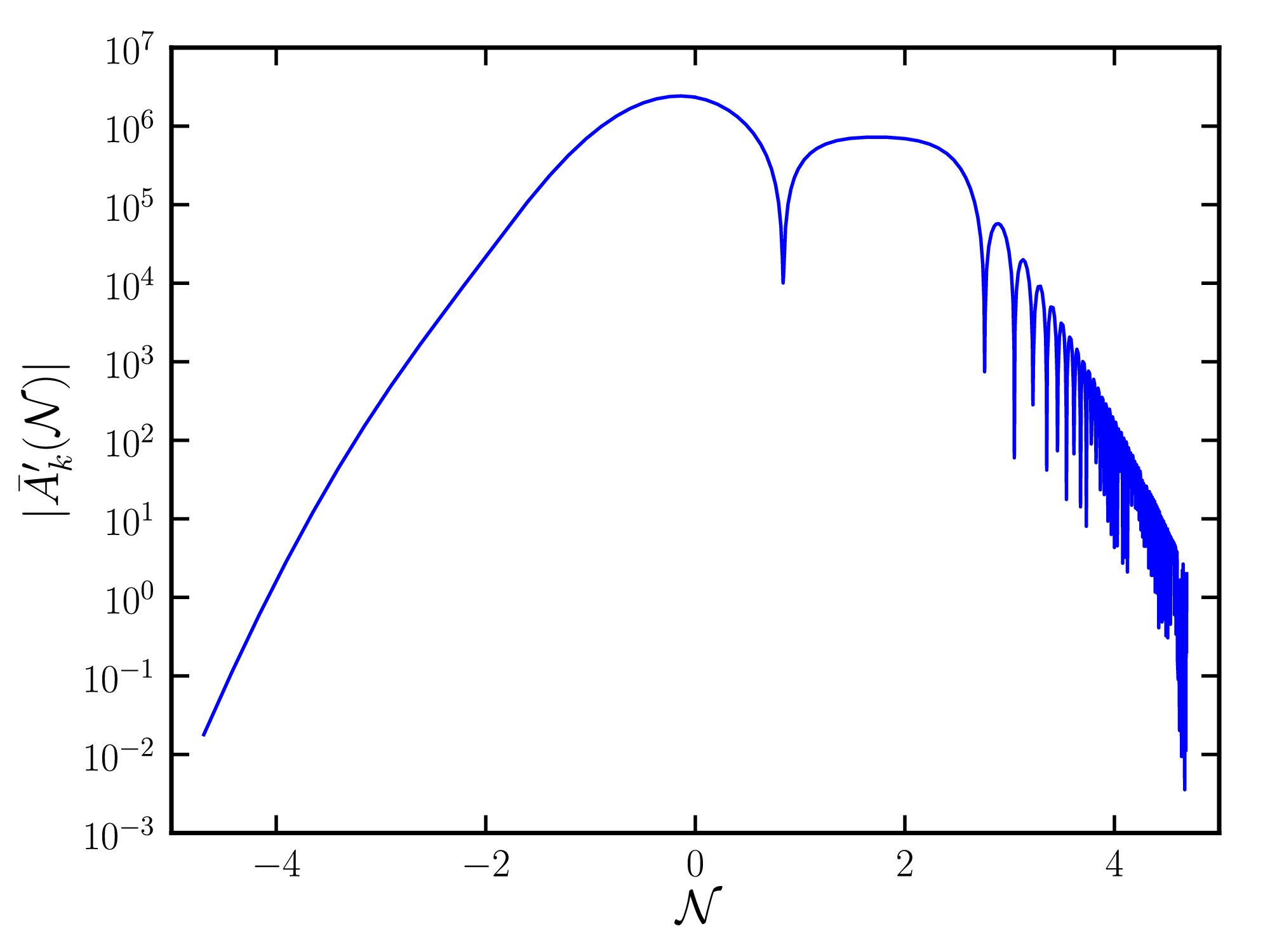}\\
\includegraphics[width=6.0cm]{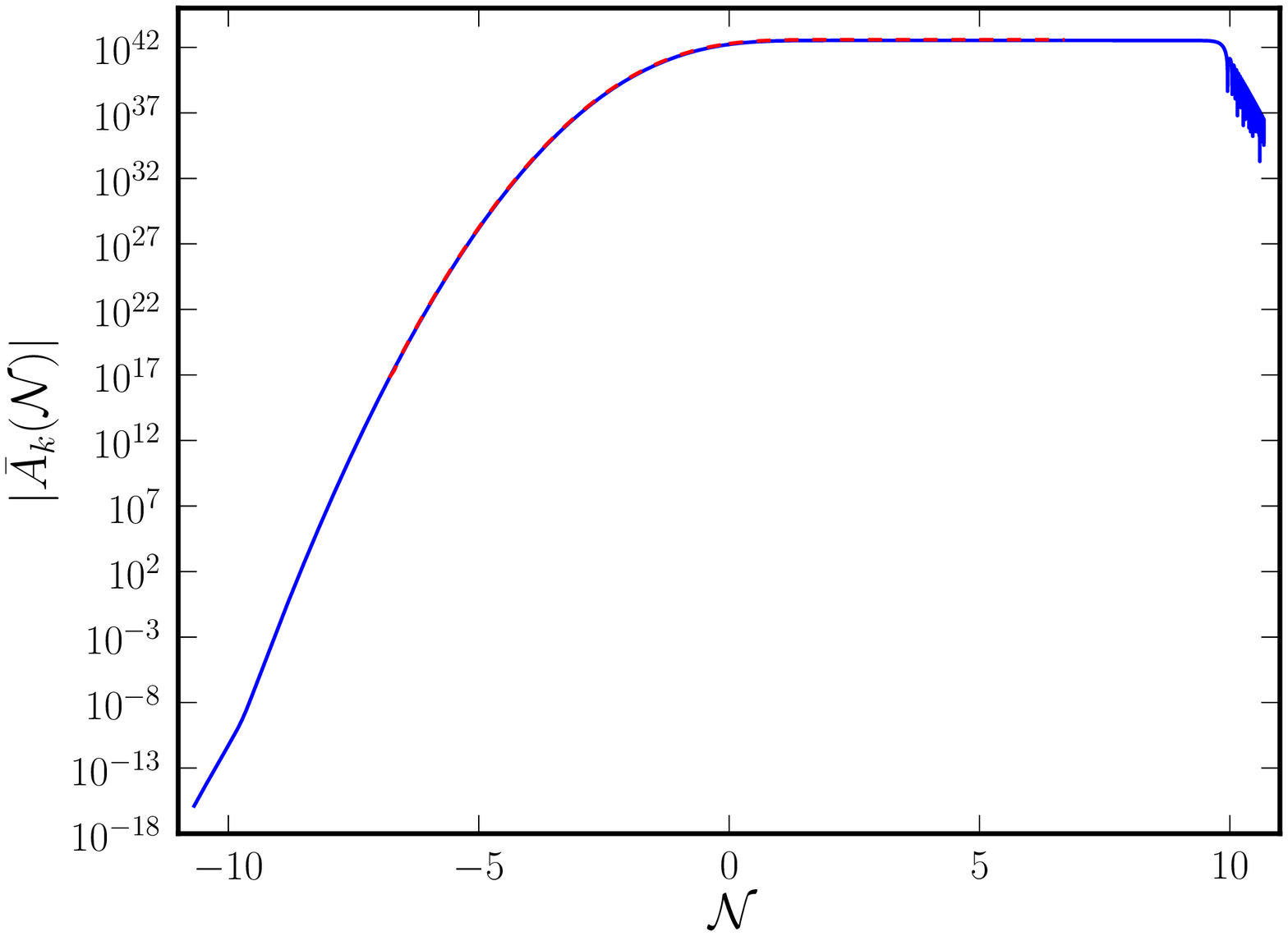}
\includegraphics[width=6.0cm]{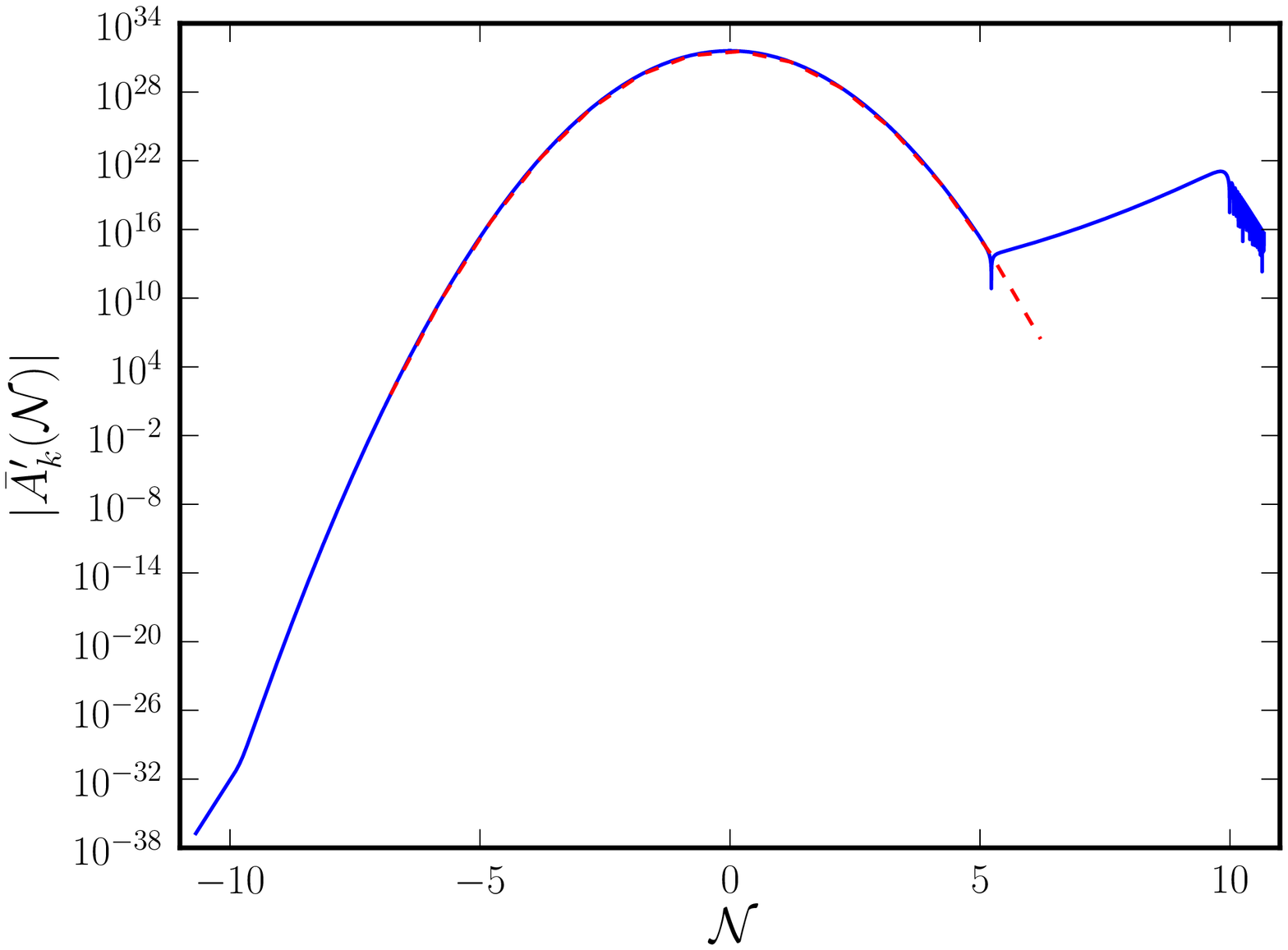}\\
\includegraphics[width=6.0cm]{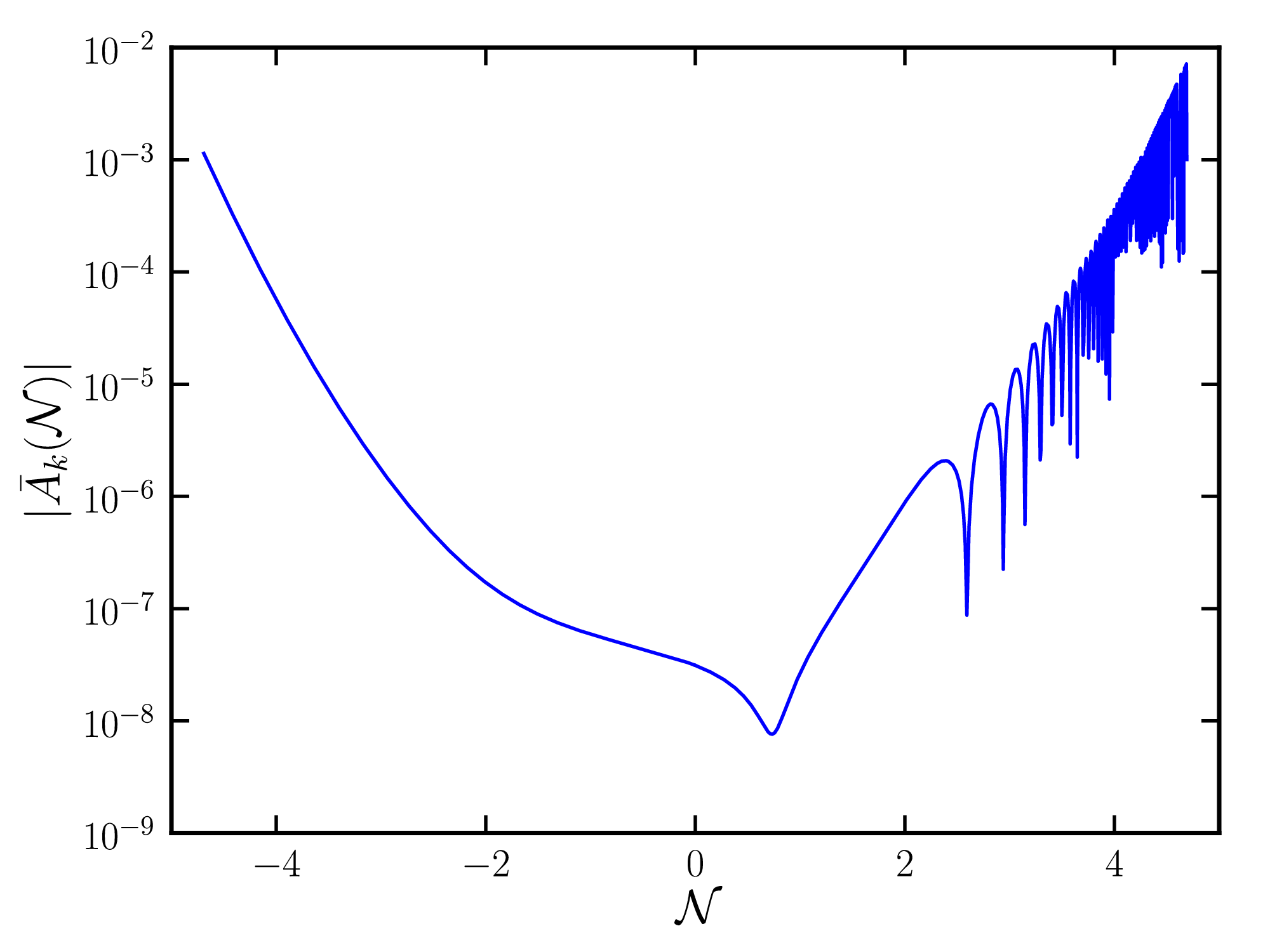}
\includegraphics[width=6.0cm]{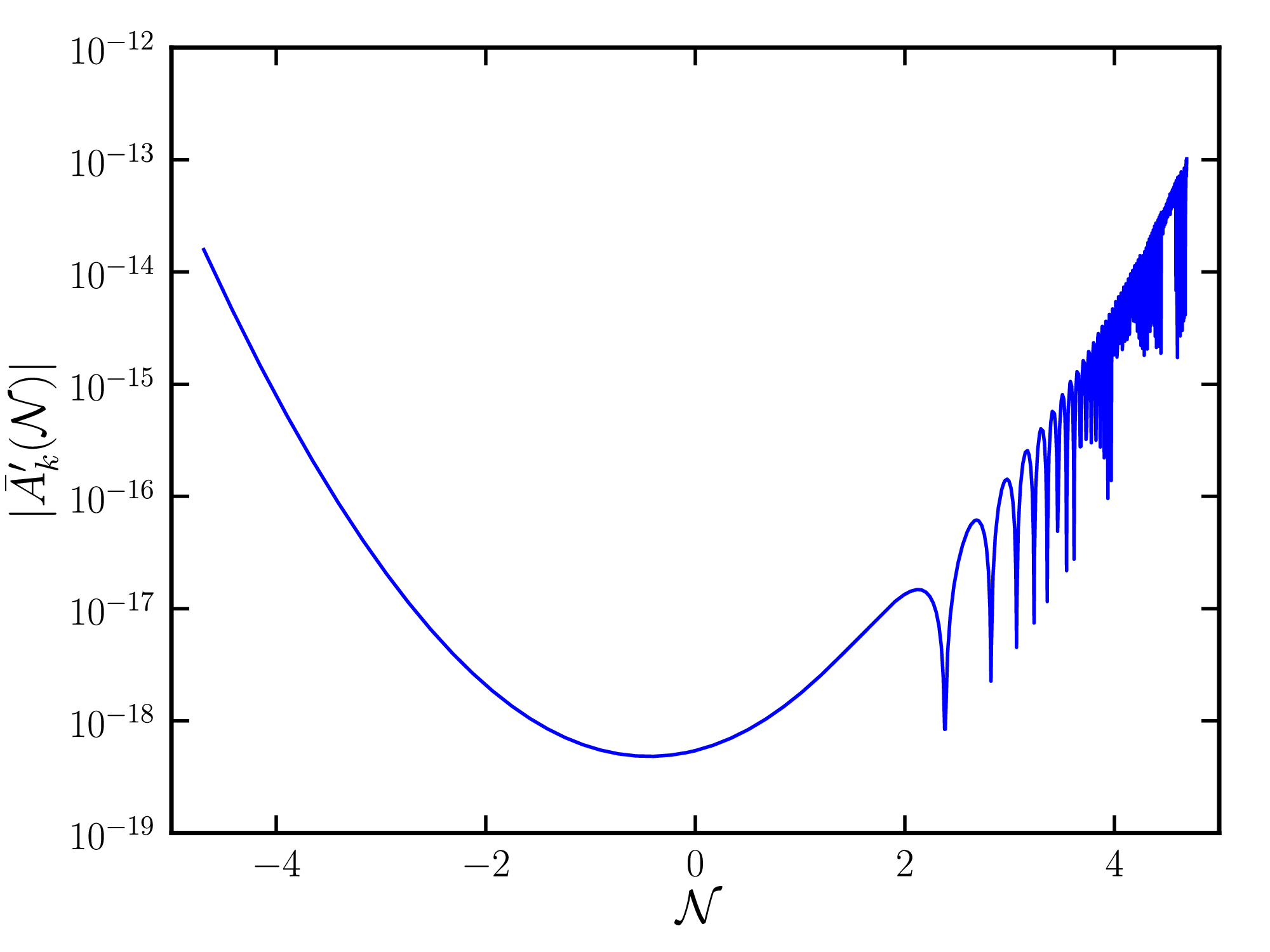}
\includegraphics[width=6.0cm]{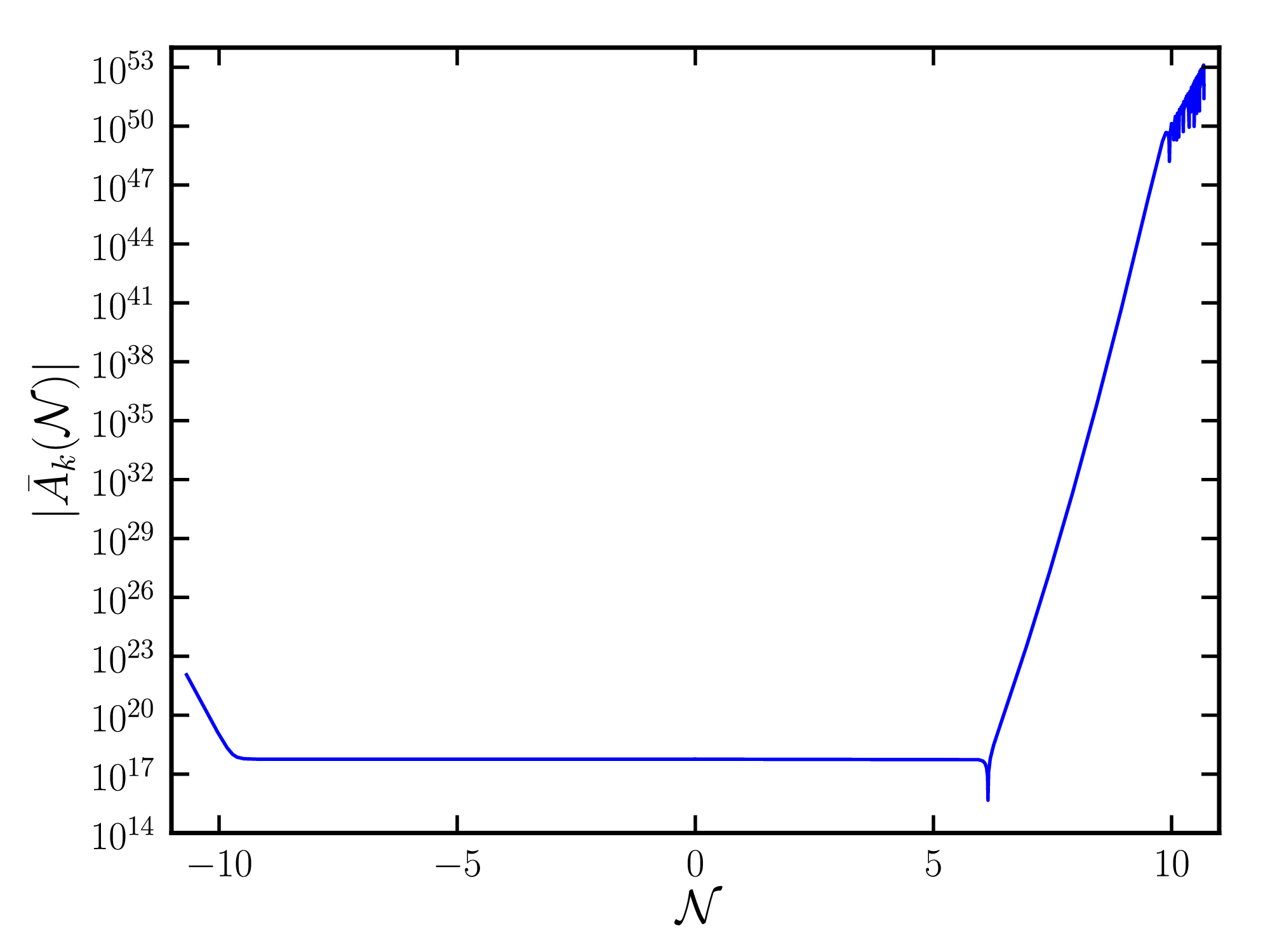}
\includegraphics[width=6.0cm]{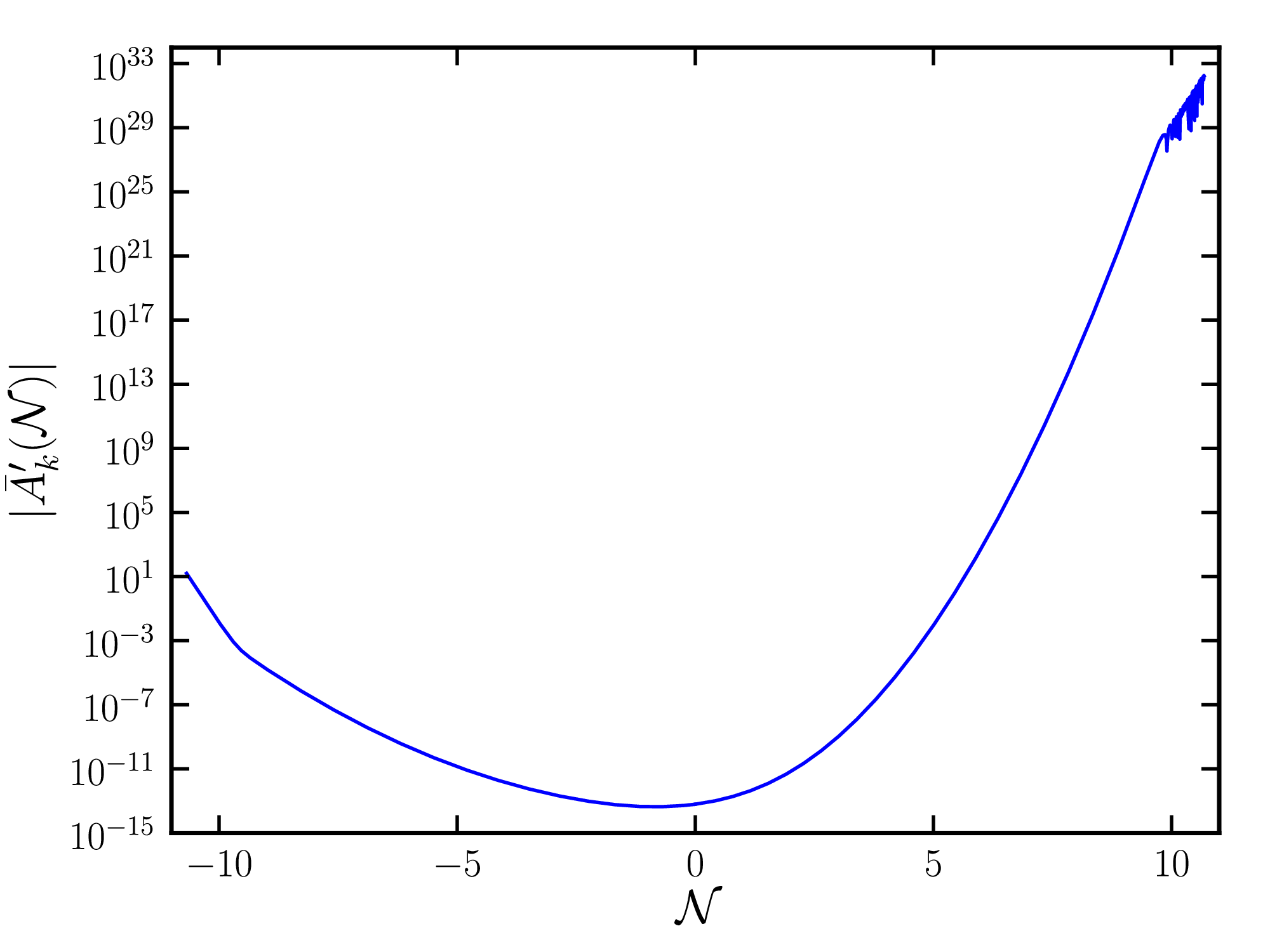}
\end{center}
\vskip -15pt
\caption{The behavior of the absolute values of $\bA_k$ (on the left) and 
its derivative $\bA_k'$ (on the right) has been plotted for the 
modes $k_0=\eta_0^{-1}={\rm e}^{-25}=1.389\times10^{-11}$ (the first and 
the third rows) and $k=10^{-10}\, k_0$ (the second and the fourth rows) for 
the cases wherein $n=3/2$ (the top two rows) and $n=-1$ (the bottom two rows).
We have set $p=1$, $a_0=10^{-10}$ and $J_0=10^4$ in all the cases.
It is interesting to note that the amplitude of $\bA_k$ has its maximum value 
at the bounce when $J$ itself has a minimum, while $\bA_k$ exhibits a minimum 
when $J$ has a maximum at the bounce.
In the absence of any detailed modeling of the bounce, we had assumed that 
$\eta_0^{-1}=\Mpl$.
So, it is indeed rather extreme to consider the mode $k_0=\eta_0^{-1}$, as scales 
of cosmological interest correspond to at least $50$ orders of magnitude smaller 
than $k_0$!
However, the evolution proves to be rather simple for much smaller wavenumbers.
We find that, for $k\ll k_0$, the absolute values of $\bA_k$ essentially grows 
(when $n$ is positive) or decays (when $n$ is negative) from its initial value
to turn to a constant as one crosses the bounce, before it decays or grows with
superposed oscillations at very late times.
The dashed red curves represent the analytical estimates~(\ref{eq:Abk-ar})
and~(\ref{eq:Abpk-ar}) and they match the numerical results as described
in the text.}
\label{fig:Ab-e}
\end{figure}

We have numerically integrated the differential equation~(\ref{eq:de-Abk})  
using a Fortran code that is based on the fifth order Runge-Kutta 
method~\cite{Press:1992zz}.
We have also independently checked the results we have obtained using 
{\sl Mathematica}.\/
In Fig.~\ref{fig:Ab-e}, we have plotted the evolution of $\bA_k$ and its time 
derivative $\bA_k'$ for two widely different modes and certain values for the 
parameters involved.
For reasons outlined above, we have worked with indices $n$ and $p$ corresponding
to $\alpha=3$ and $-2$.
It is useful to note from the figures that, when $J$ has a minimum at the 
bounce (\ie\/ when $n$ is positive), ${\bar A}_k$ exhibits a maximum, 
whereas ${\bar A}_k$ has a minimum at the bounce when $J$ has a maximum 
(\ie\/ when $n$ is negative). 
Moreover, it is clear from the figures that, for $k\ll k_0$, there 
exists a wide domain in time over which ${\bar A}_k$ proves to be 
a constant.
It is over this domain that we shall choose to evaluate the power spectra
of the magnetic and the electric fields.
In the figure corresponding to $p=1$ and $n=3/2$ (leading to $\alpha=3$) and 
$k\ll k_0$, we have also illustrated the analytical results~(\ref{eq:Abk-ar}) 
and~(\ref{eq:Abpk-ar}).
It is clear that the analytical result for $\bA_k'$ matches the numerical
result very well.
The behavior of $\bA_k$ is also along expected lines.
Around the bounce, the analytical solution~(\ref{eq:Abk-ar}) we have obtained
mimics the exact numerical result extremely well.

\par

\begin{figure}[!h]
\begin{center}
\includegraphics[width=7.6cm]{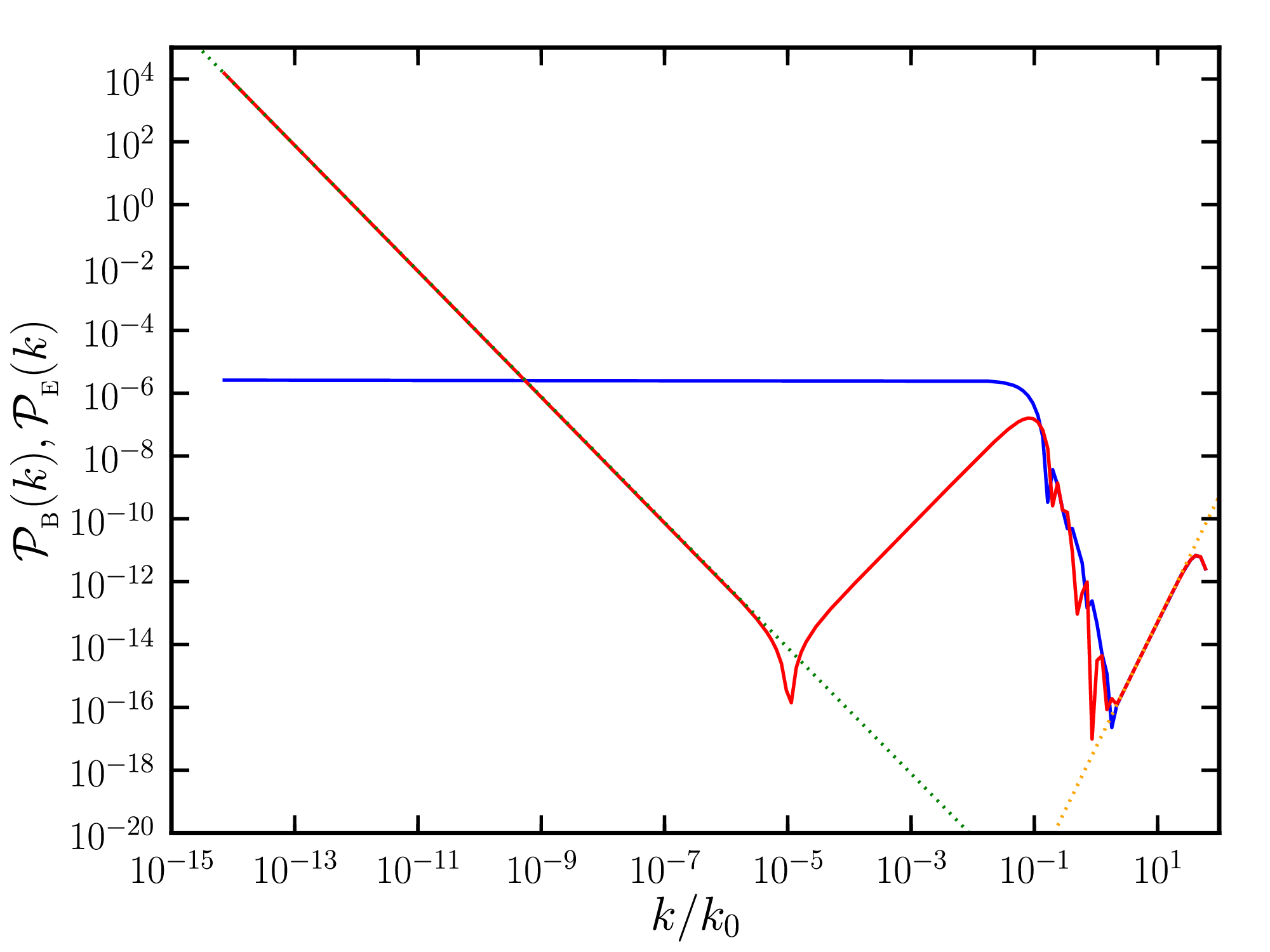}
\includegraphics[width=7.6cm]{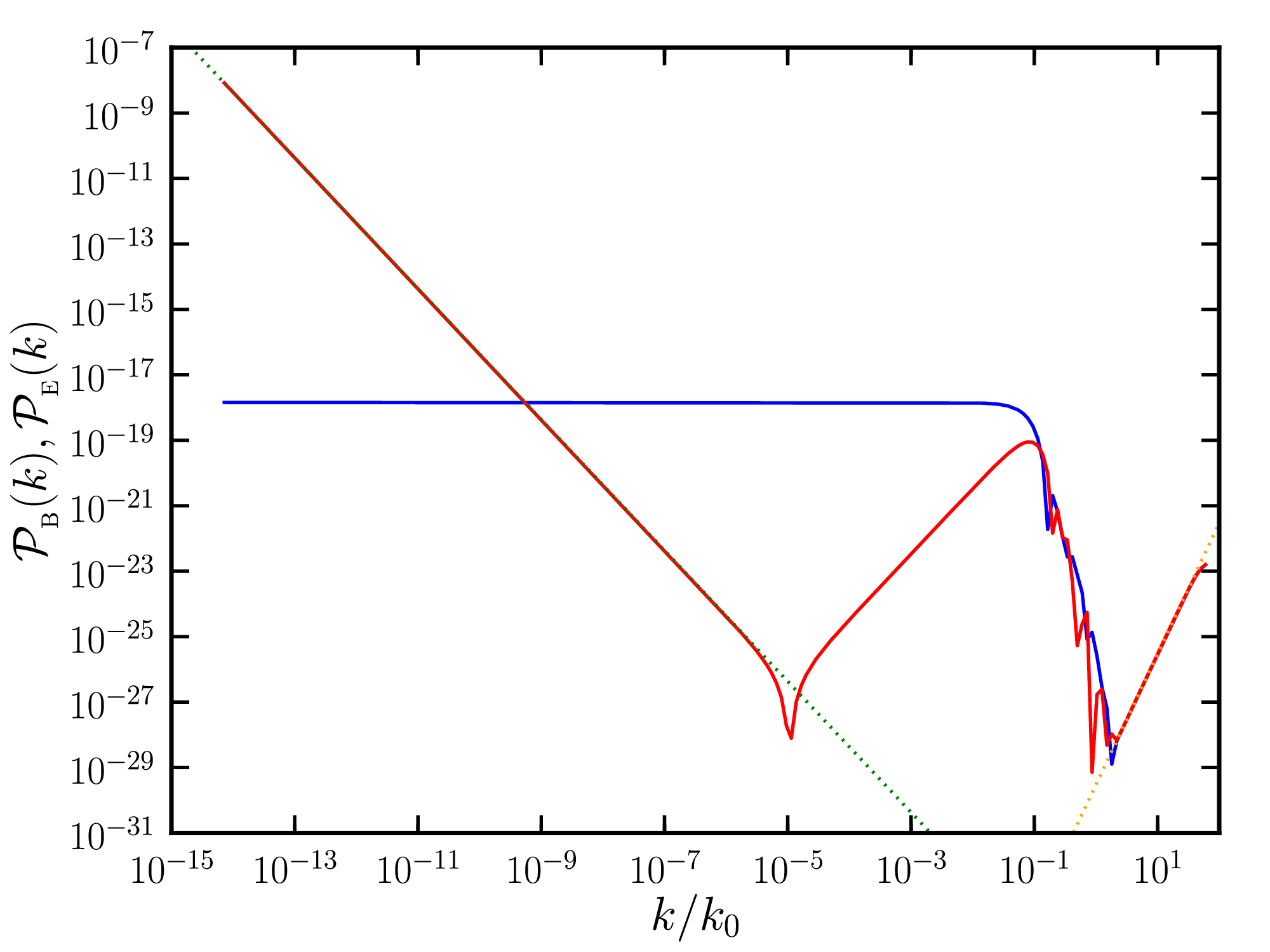}
\includegraphics[width=7.6cm]{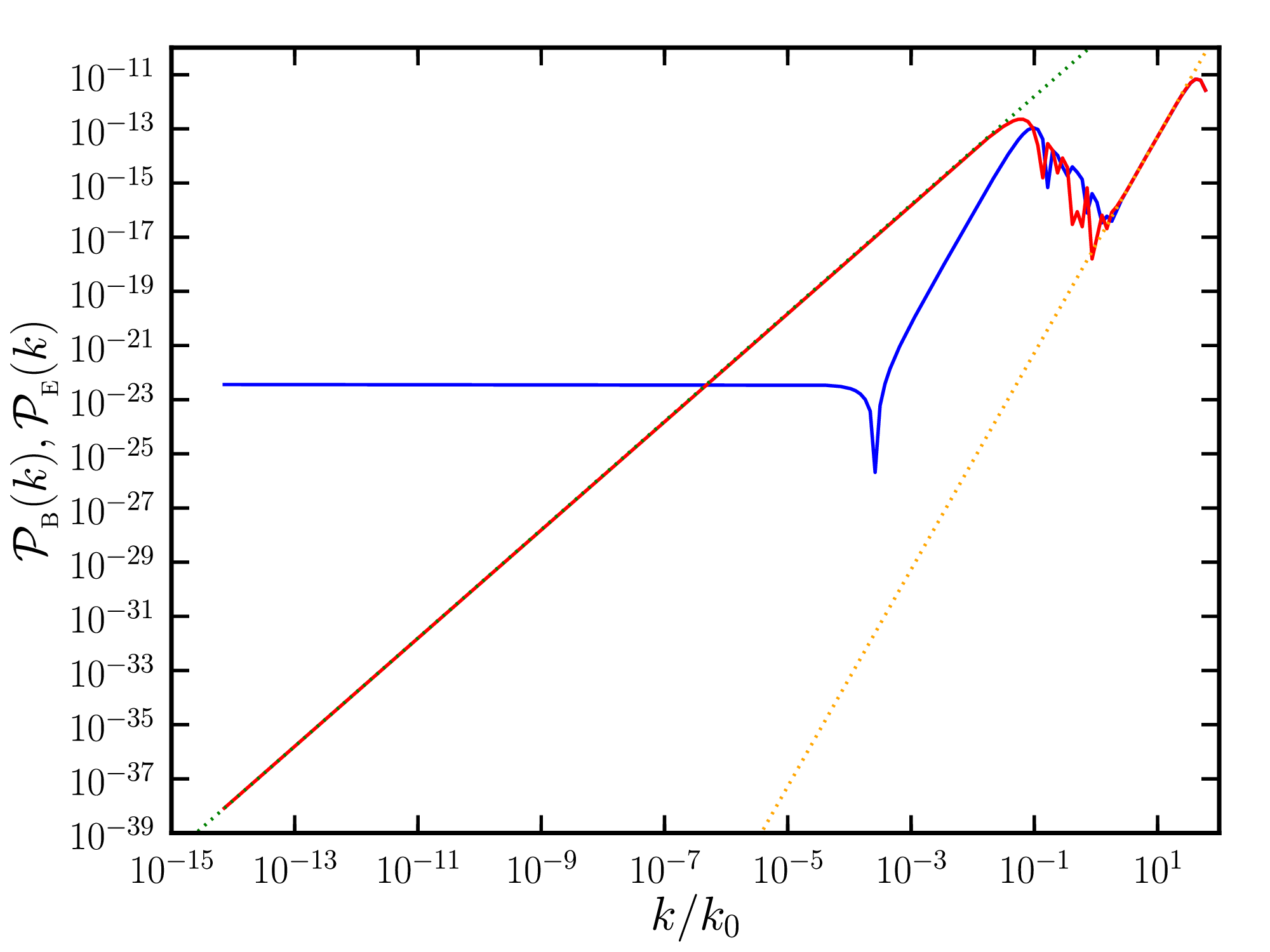}
\includegraphics[width=7.6cm]{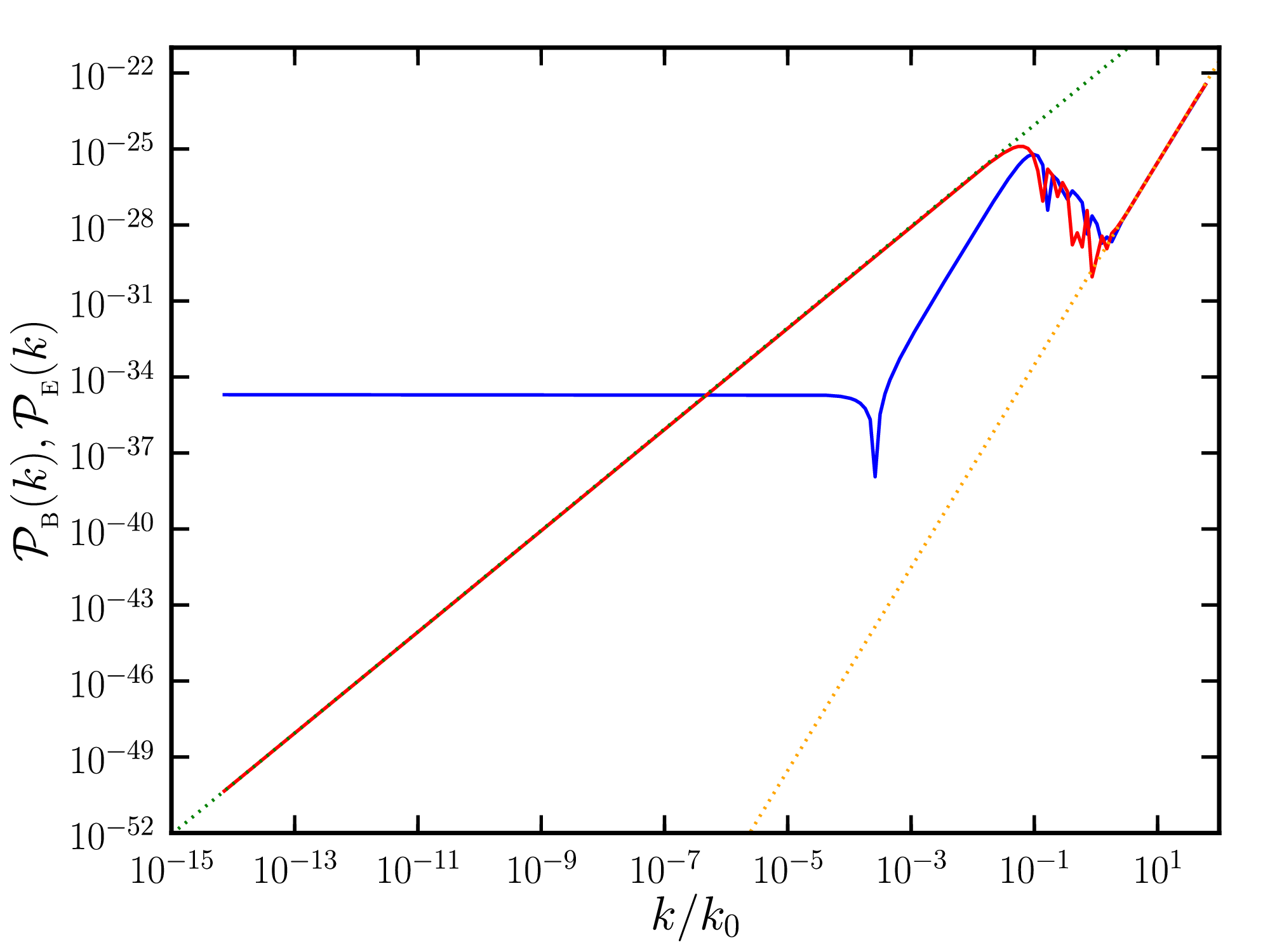}
\end{center}
\caption{The power spectra of the magnetic (in blue) and the electric (in 
red) fields for the cases wherein $(p,n)=(1,3/2)$ (top left), $(p,n)=(2,3/4)$ 
(top right), $(p,n)=(1,-1)$ (bottom left) and $(p,n)=(2,-1/2)$ (bottom right).
Evidently, the top and the bottom rows correspond to $\alpha=3$ and $\alpha=-2$, 
respectively.
We should stress that we have evaluated these spectra {\it after}\/ the bounce.
We have worked with the same values of $\eta_0$, $a_0$ and $J_0$, \viz\/ 
$\eta_0={\rm e}^{25}=7.2\times10^{10}$, $a_0=10^{-10}$ and $J_0=10^4$, 
in all the four cases.
As expected from the analytical arguments, the spectra for a given $\alpha
=2\,n\,p$ (\ie\/ within each row) have the same shape.
However, note that, for a given $\alpha$, the amplitude differs for different
$p$ and $n$.
Clearly, the spectrum of the magnetic field is scale invariant in all these 
cases.
The power spectra of the electric field are along expected lines, behaving as
$k^{4-2\,\alpha}=k^{-2}$ when $\alpha=3$ and $k^{6+2\,\alpha}=k^2$ when 
$\alpha=-2$ (indicated by the dotted green lines).
For wavenumbers much smaller than the bounce scale $k_0$, the power in the 
electric field is smaller than the power in the magnetic field when $\alpha=-2$, 
\ie\/ when $n$ is negative.
Note that, modes with wavenumbers comparable to or greater than the bounce scale 
$k_0$ are hardly affected by the bounce.
Hence, over these scales, the power spectra of both the electric and magnetic 
fields are expected to behave in a similar manner (in fact, as $k^4$, indicated
by the dotted orange lines), which is exactly the dependence that we obtain 
from the numerical results.}
\label{fig:ps}
\end{figure}

Let us now turn to the evaluation of the power spectra of the magnetic and 
and electric fields generated in the cases of interest.
Let us first list out all the parameters that we have.
These parameters appear in the functions describing the scale factor $a$ and 
the non-minimal coupling $J$.
The parameters that characterize the scale factor are $a_0$, $\eta_0$ and $p$.
Apart from these three, two additional parameters, \viz\/ $J_0$ and $n$, are
required to describe the coupling function $J$. 
As we have already discussed, $k_0=\eta_0^{-1}$ determines the scale associated
with the bounce.
We shall plot the power spectra in terms of $k/k_0$.
Clearly, our first goal would be to examine if we obtain scale invariant spectra 
for the magnetic field for any values of the parameters.
As we have already discussed, for $k\ll k_0$, we expect scale invariant spectra 
for the magnetic field before the bounce when $\alpha=3$ or $-2$.
Also, based on the same arguments, one can show that, before the bounce, we can 
expect the power spectrum of the electric field behave as $k^{4-2\,\alpha}$ when 
$\alpha=3$ and as $k^{6+2\,\alpha}$ when $\alpha=-2$.
With the numerical tools at hand, it is interesting to examine if these power 
spectra retain their shape after the bounce as well. 
In Fig.~\ref{fig:ps}, we have plotted the power spectra $\psb(k)$ and $\pse(k$)
for a set of cases that lead to scale invariant spectra for the magnetic field.
It is clear from the figure that the spectra indeed retain the shape for small
wavenumbers (expected from the analytical arguments) as they emerge through 
the bounce. 
An interesting point to note from these figures is the behavior of the spectra
for $k\gtrsim k_0$.
These modes are hardly affected by the background and they retain their original
form determined by the initial conditions.
As a result, the Minkowski-like initial conditions that we have imposed imply 
that both the magnetic and electric fields should behave as $k^4$.
This is exactly the behavior that we obtain from the numerical results.
The fact that the analytically expected results are reproduced indicates 
the robustness of the numerical procedures that have been adopted.   

\par

Note that the definitions of the power spectra~(\ref{eq:ps}) contain an 
overall factor of $J^2$. 
Further, the differential equation~(\ref{eq:de-Abk}) satisfied by $\bA_k$
only involves the ratio $(\d J/\d{\cal N})\,(1/J)$. 
Moreover, note that the initial conditions~(\ref{eq:ic}) contain a factor
of $J$ in the denominator and
\clearpage\noindent
will hence will involve a factor of $1/J_0$.
Therefore, the power spectra are independent of $J_0$.
(In fact, $J$ also contains the parameter $a_0$. 
But, it can be absorbed in the overall constant $J_0$ leading to the same
conclusions as above.)
This can also be easily confirmed with the numerics. 
The amplitude of the power spectra are determined by the parameters $a_0$ 
and $\eta_0$ (apart from the indices $p$ and $n$ which also influence
the amplitude).
Note that, since it is the combination $k/a_0$ that appears in the differential 
equation governing $\bA_k$, $a_0$ simply sets the scale.
Therefore, the dominant dependence of the power spectra on $a_0$ arises due
to the factor of $a^4$ that appear in the denominators of the power spectra
[cf. Eqs.~(\ref{eq:ps})].
Hence, we can expect the power spectra to behave as $a_0^{-4}$. 
For a given set of the other parameters, we find numerically that the amplitudes 
of the spectra indeed behave in such a fashion.


\section{Generating magnetic fields of observable strengths}\label{sec:os}

We have established that scale invariant spectra of magnetic fields can 
be generated in bouncing models.
Let us now examine if the strengths of these primordial magnetic fields 
can correspond to observable levels today.

\par

\begin{figure}[!t]
\begin{center}
\includegraphics[width=15.0cm]{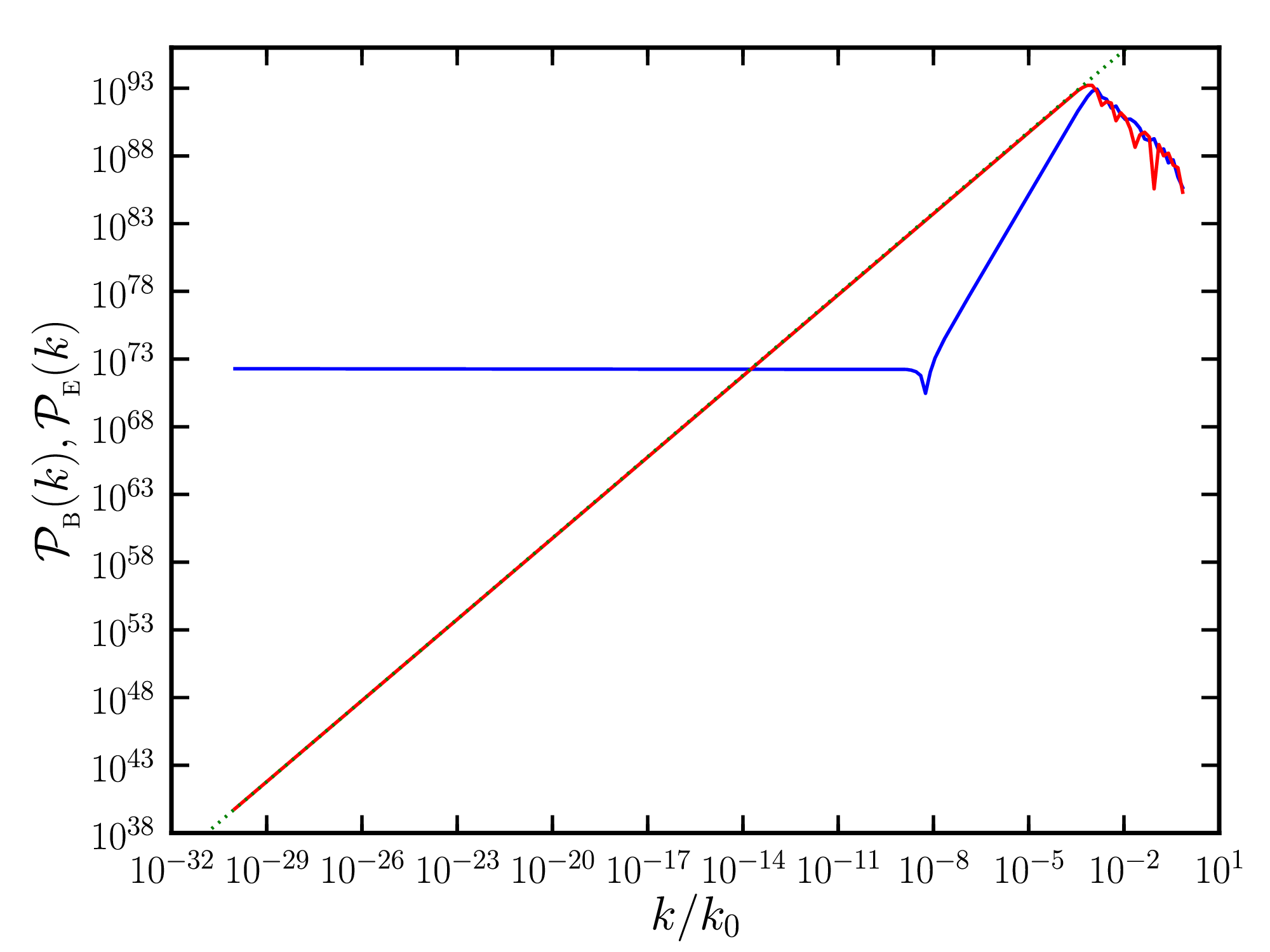}
\end{center}
\caption{The power spectra with $p=1$ and $n=-1$, corresponding to $\alpha=-2$
has been plotted for a wide range of wavenumbers (with same choice of colors to
represent the spectra as in the previous figure).
We have set $\eta_0=1$, $a_0=4\times10^{-29}$ and $J_0=10^4$ in arriving at these
results.
As we have discussed, the amplitude of the spectra are independent of the 
choice of $J_0$, which we also observe numerically.
Clearly, a suitably small value of $a_0$ leads to magnetic fields in the early 
universe that correspond to observable strengths today.
Note that we have worked with parameters such that $\alpha=-2$ as it results
in electric fields of strengths considerably smaller than that of the magnetic
fields over observationally relevant scales.}
\label{fig:ps-fv}
\end{figure}

As the issue of reheating in bouncing models remains poorly understood, 
we shall simply assume that, at some stage after the bounce, the universe 
transits to the radiation dominated epoch. 
We shall also assume that the coupling function $J$ simultaneously reduces 
to unity.
Let the bouncing phase end and the radiation dominated epoch begins at the 
temperature, say, $T_{\rm end}$, corresponding to the scale factor, say,
$a_{\rm end}$.
Under these conditions, the power spectrum of the magnetic field at 
the epoch corresponding to $T_{\rm end}$, say, $\psb^{\rm end}(k)$, is 
related to the power spectrum observed today, say, $\psb^{\rm today}(k)$, 
through the relation  
\begin{equation}
\psb^{\rm end}(k)
= \psb^{\rm today}(k)\, \l(\f{a_{\rm today}}{a_{\rm end}}\r)^4
=\psb^{\rm today}(k)\, \l(\f{T_{\rm end}}{T_{\rm today}}\r)^4,
\end{equation}
where $T_{\rm today}$ is the temperature today.
If we assume that $T_{\rm end}\simeq 10^{13}\,{\rm GeV}$ (a choice that
will be explained below) and, since $T_{\rm today}\simeq 10^{-4}\, {\rm eV}$, 
in order to generate magnetic fields of observed strengths today, \ie\/ 
$B_{\rm today}\simeq 10^{-16}\, {\rm Gauss}$ (in this context, see, for example,
Refs.~\cite{Neronov:1900zz,Tavecchio:2010mk,Yamazaki:2010nf,Kahniashvili:2010wm,Caprini:2011cw,Ade:2015cva}),
the power spectrum $\psb^{\rm end}(k)$ should be of the order of
\begin{equation}
\psb^{\rm end}(k)
=\l(10^{-16}\r)^2\, \l(\f{10^{22}\, {\rm eV}}{10^{-4}\, {\rm eV}}\r)^4\, 
{\rm Gauss}^2
\simeq 10^{72}\, {\rm Gauss}^2.
\end{equation}
We had discussed earlier that the power spectra behave as $a_0^{-4}$.
We find that it is indeed possible to produce magnetic fields of such
large strengths by working with a suitably small value of $a_0$.    
In Fig.~\ref{fig:ps-fv}, we have plotted the power spectra with the
required strength for certain values of the parameters.  
A couple of related points require clarification.
We had mentioned earlier that the spectra are evaluated at a given time when 
the smallest scale of interest satisfies the condition $k^2=10^3\, (J''/J)$.
In plotting Fig.~\ref{fig:ps-fv}, we have chosen the smallest scale to be 
$k_0$.
For $\eta_0=1$, $p=1$ and $n=-1$, we find that this corresponds to evaluating 
the spectra at the ${\rm e}$-${\cal N}$-fold of roughly ${\cal N}=5.5$.
This suggests that we can choose $a_{\rm end}\simeq \exp\,(5.5^2/2)\, a_0
\simeq 10^{6}\,a_0$.
If the bounce corresponds to the Planck scale $\Mpl$, then $T_{\rm end}$ can 
be chosen to be $T_{\rm end}=\Mpl\, (a_0/a_{\rm end})=\Mpl/10^6\simeq
10^{13}\, {\rm GeV}$.
It is this choice that we have made above.
We should add that a more complete calculation relating the strengths of the 
magnetic fields soon after the bounce and the observed strengths today will
require a good understanding of the transition from the bounce to the epoch
of radiation domination.


\section{The issue of backreaction}\label{sec:br}

In this section, we shall discuss the issue of backreaction in the bouncing 
models of our interest.
In the inflationary context, as we have discussed, scale invariant magnetic 
fields are generated when the coupling function $J$ either grows with the 
scale factor or decays in certain manner. 
While the former case suffers from the strong coupling 
problem~\cite{Demozzi:2009fu,Ferreira:2013sqa,Ferreira:2014hma}, the issue
of backreaction becomes important in the latter.
In particular, in the latter case, the energy density associated with the 
generated electromagnetic fields rapidly grow with time and can dominate 
the energy density that has been assumed to drive the background evolution 
(see, for instance, Refs.~\cite{Kanno:2009ei,Urban:2011bu}).
Such a behavior is untenable and, for the scenario to remain viable, the energy 
densities in the electromagnetic fields that have been generated should always 
remain sub-dominant to the energy density associated with the background.
Let us now examine if this condition is satisfied in the scenarios that we
have considered here.

\par

Recall that, from the Friedmann equation, we have 
\begin{equation}
\rho_{\rm bg}=3\, \Mpl^2\, H^2= 3\, \Mpl^2\,\l(\f{a'}{a^2}\r)^2,
\end{equation}
where $\rho_{\rm bg}$ is the energy density that is driving the background
evolution.
Given the scale factor~(\ref{eq:sf}), the corresponding energy density can
be expressed as
\begin{equation}
\rho_{\rm bg}=\f{12\,p^2 \Mpl^2}{a_0^2\,\eta_0^2}\,
\l[\l(\f{a}{a_0}\r)^{1/p}-1\r]\,\l(\f{a}{a_0}\r)^{-2\,(p+1)/p}.
\end{equation}
The energy density in a specific mode $k$ of the electromagnetic field is 
given by
\begin{equation}
\rho_{_{\rm EB}}^k=\psb(k)+\pse(k). 
\end{equation}
Evidently, if the backreaction is to be negligible, we require that
$\rho_{\rm bg}>\rho_{_{\rm EB}}^k$ for all modes of cosmological interest.
Also, this condition should hold true at all times.
However, note that, since $H$ vanishes at the bounce, $\rho_{\rm bg}$ 
does so too.
A priori, it should be clear that any non-trivial amount of energy density 
in the electromagnetic fields that have been generated will lead to a 
violation of the required condition.

\par

\begin{figure}[!t]
\begin{center}
\includegraphics[width=15.0cm]{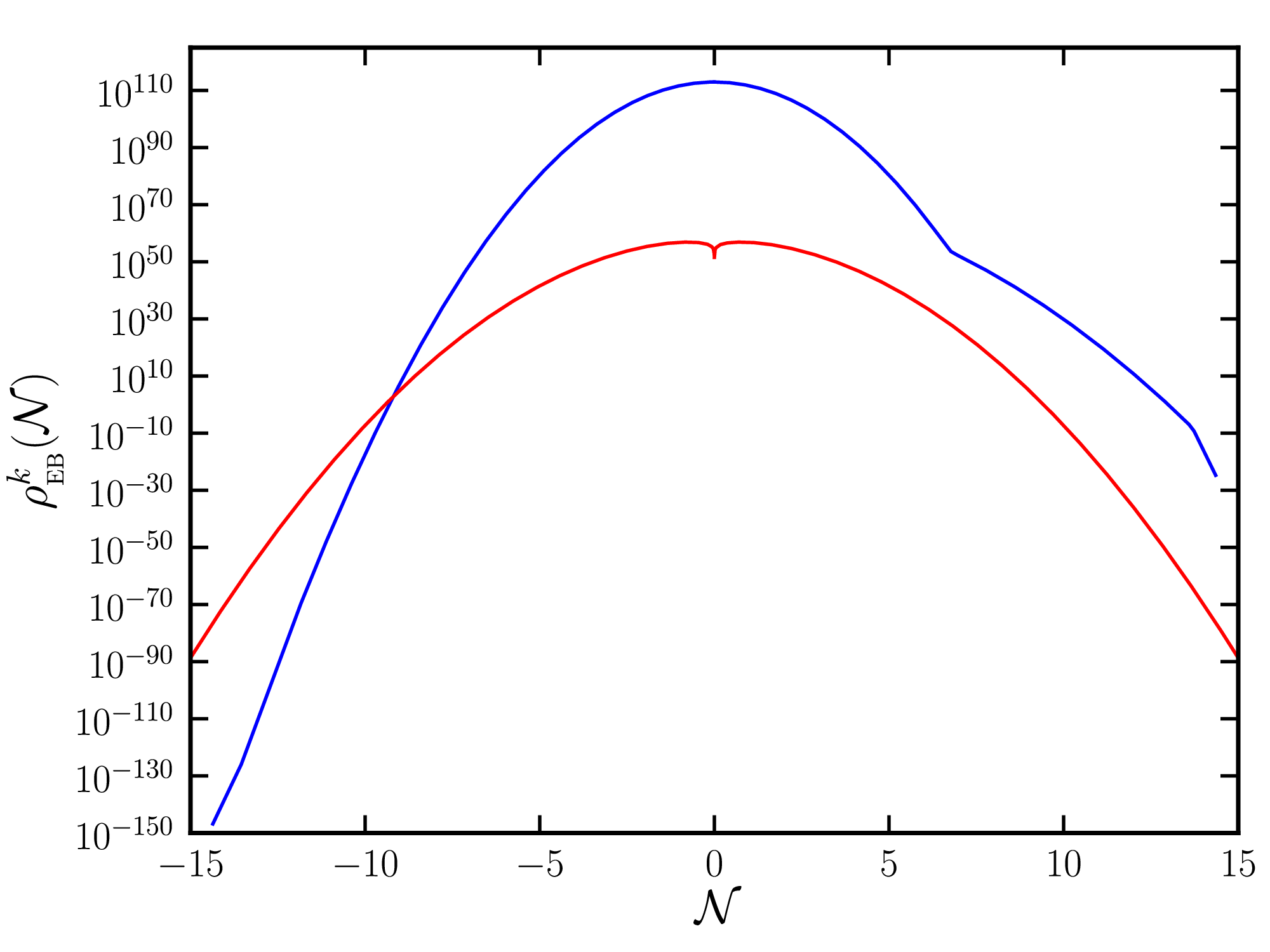}
\end{center}
\caption{The behavior of the energy density in the electric and magnetic
fields for the mode $k=10^{-20}\, k_0$ has been plotted (in blue) along 
with the energy density of the background (in red).
We have worked with the same values of the various parameters as in the 
last figure.
In the previous figure, we had arrived at magnetic fields that correspond 
to observable strengths today by choosing to work with an extremely small 
value of $a_0$.  
Such a small value, unfortunately, boosts the energy density in the 
generated electromagnetic fields. 
The energy density in the electromagnetic fields proves to be unreasonably
large at the bounce clearly calling into question the viability of the
model.}
\label{fig:br}
\end{figure}

Let us nevertheless estimate the energy density associated with the
electromagnetic modes that have been created.
The dependence of $\rho_{_{\rm EB}}^k$ on time for a particular mode can 
be easily arrived at from the numerical solutions we have obtained.
In Fig.~\ref{fig:br}, we have plotted $\rho_{_{\rm EB}}^k$ for a relatively
large scale mode, along with the background energy density $\rho_{\rm bg}$.
We have plotted the results for values of the parameters that we had 
considered in arriving at magnetic fields of observable strengths in
Fig.~\ref{fig:ps-fv}.
It is clear from Fig.~~\ref{fig:br} that the energy density in the electromagnetic 
field (of the given mode) is smaller than the energy density of the background at 
early stages of the bounce.
However, as one approaches the bounce, the energy density of the electromagnetic 
field grows rather quickly beyond the background energy density, leading to a 
severe violation of the expected condition.
Needless to add, this issue of backreaction has to be circumvented if bouncing models 
are considered to be a viable scenario for the generation of observable levels
of magnetic fields.


\section{Discussion}\label{sec:d}

In this work, we have studied the generation of primordial magnetic fields in a class 
of bouncing universes when the electromagnetic field is coupled non-minimally 
to a scalar field that drives the background expansion. 
We had restricted ourselves to the consideration of symmetric non-singular bouncing 
models that allow initial conditions on the perturbations to be imposed at sub-Hubble 
scales at very early times during the contracting phase of the universe. 
We found that there exists a class of indices describing the non-minimal coupling 
and the scale factor that lead to a nearly scale invariant spectrum for the
magnetic field, while the corresponding electric field spectrum is sharply
scale dependent.
We showed that certain values of the parameters involved lead to primordial
magnetic fields which correspond to observable strengths today.
However, unfortunately, the backreaction due to the electromagnetic fields
that have been generated prove to be substantial calling into question the
viability of the model.

\par

Although we have not discussed how to obtain the desired bouncing solution, it turns out 
that bouncing scenarios typically require a violation of the null energy condition at the 
bounce. While this is true for spatially flat or open FLRW universes, the necessity to 
violate the null energy condition can be circumvented with a positive spatial curvature. 
Since the backreaction due to the generated electromagnetic fields is an issue at the 
bounce in spatially flat models, an interesting and possible way to avoid it would be to 
consider models with a positive but small spatial curvature. Such a spatial curvature 
would require a non-vanishing energy density at the bounce and hence may aid in 
overcoming the backreaction problem. We are currently exploring such issues.

\acknowledgments

We would like to thank J\'er\^ome Martin, T.~R.~Seshadri and Kandaswamy 
Subramanian for useful discussions and comments on the manuscript.
We also wish to thank Agustin Membiela for detailed comments on the manuscript.
LS wishes to thank the Indian Institute of Technology Madras, Chennai, 
India, for support through the New Faculty Seed Grant.
RKJ acknowledges financial support  from the Danish council for independent 
research in Natural Sciences for part of the work. 

\bibliographystyle{JHEP}
\bibliography{gmf-bu-refs}

\end{document}